\documentclass[floatfix,prb,epsf,twocolumn,superscriptaddress,eqsecnum]{revtex4}
\usepackage{amsfonts}
\usepackage{bbold}
\usepackage{amsmath,amssymb,natbib,bm,graphicx,url,epsfig}
\usepackage{bm}
\usepackage{bbm}
\usepackage[ansinew]{inputenc}
\usepackage[active]{srcltx}
\usepackage[usenames, dvipsnames]{color}
\usepackage{dsfont}

\newcommand{\be}{\begin{equation}}
\newcommand{\ee}{\end{equation}}
\newcommand{\bes}{\begin{equation}\begin{split}}
\newcommand{\ees}{\end{split}\end{equation}}
\newcommand{\bea}{\begin{eqnarray}}
\newcommand{\ena}{\end{eqnarray}}
\newcommand{\nn}{\nonumber}

\begin{document}

\title{Supersymmetry method for interacting chaotic and disordered systems:\\
the SYK model}
\author{Tigran A. Sedrakyan}
\affiliation{Department of Physics, University of Massachusetts, Amherst, Massachusetts
01003, USA}
\author{Konstantin B. Efetov}
\affiliation{Ruhr University Bochum, Faculty of Physics and Astronomy, Bochum, 44780,
Germany }
\affiliation{National University of Science and Technology \textquotedblleft
MISiS\textquotedblright , Moscow, 119049, Russia}

\begin{abstract}
The nonlinear supermatrix $\sigma $-model is widely used to understand the physics of Anderson localization and the level statistics in noninteracting disordered electron systems. In contrast to the general belief that the supersymmetry method applies only to systems of noninteracting particles, we adopt this approach to the disorder averaging in the interacting models. In particular, we apply supersymmetry to study the Sachdev-Ye-Kitaev (SYK) model, where the disorder averaging has so far been performed only within the replica approach. We use a slightly modified, time-reversal invariant version of the SYK model and perform calculations in real-time. As a demonstration of how the supersymmetry method works, we derive saddle point equations. In the semiclassical limit, we show that the results are in agreement with those found using the replica technique. We also develop the formally exact superbosonized representation of the SYK model. In the latter, the supersymmetric theory of original fermions and their superpartner bosons is reformulated as a model of unconstrained
collective excitations. We argue that the supersymmetry description
of the model paves the way for precise calculations in SYK-like
models used in condensed matter, gravity, and high energy physics.
\end{abstract}

\maketitle


\section{Introduction}

The study of disordered and chaotic systems is a prevalent topic in
condensed matter physics, and various models of interacting particles have been under intensive investigation for more than half a century. Less
expected has been a recent application of models with disorder to gravity
and quantum field theory \cite{Kitaev-2015,Sachdev-2015,Maldacena-2016}. This field of research is fast-growing, and the study of disorder and chaos can nowadays be considered as an interdisciplinary. The latter, in particular, means that methods of calculations developed in condensed matter theory can be used in gravitation and high energy physics. 

Of course, one can simply use diagrammatic expansions in both the
interaction and disorder \cite{agd} and sum the most important diagrams, as it has been done in Ref. \cite{Maldacena-2016}. However, this approximation does not generally give full information about the system, and one has to use non-perturbative methods. 

Quantum phenomena in disordered or chaotic systems can efficiently be
investigated analytically using methods of quantum field theory. Three most
popular approaches are based on the replica trick\cite{EA}, the Keldysh
technique\cite{LVK,Schwinger,Feynman,AK}, and the supersymmetric $\sigma $%
-model approach originally developed by one of the authors\cite%
{Efetov-1983,Efetov-1997}. The necessity of applying these techniques stems
from the fact that physical correlation functions of interest are expressed
in terms of functional integrals containing weight denominators while
averaging over quenched disorder has to be done at the end of calculations.
This makes a direct application of methods of quantum field theory
difficult. All the methods of Refs. %
\onlinecite{EA,LVK,Schwinger,Feynman,AK,Efetov-1983,Efetov-1997,finkelstein}
allow one to eliminate the weight denominator $Z$ -- the partition function
of the system -- and average over disorder just at the beginning of all
calculations. As a result of this manipulation, one obtains an effective
field theory for \textquotedblleft interacting\textquotedblright\ particles
and application of well-developed methods and approximations become feasible.

Although the replica, Keldysh, and supersymmetry techniques look similar to
each other, their efficiency when applying to different problems is very
different. The replica approach allows one to avoid explicitly calculating $%
Z $ by introducing an integer number of copies of the system and making use
of the replica trick. It can be used for various systems of interacting
particles, spins, etc., but the method requires an analytical continuation
to non-integer numbers of replicas and assumes the existence of the replica
limit when the number of copies $n$ $\rightarrow 0$. A general procedure of
this continuation does not exist, and one obtains very often unphysical
results in certain situations, although one can also obtain important
results using this method \cite{finkelstein}. Within the Keldysh technique,
one doubles the degrees of freedom to obtain a normalized theory with
partition function, $Z=1$. The Keldysh sigma model representation of
disordered systems is formally exact, but it can be quite complicated for
some specific cases. Both approaches have been successfully applied to
interacting theories with the disorder, but their efficiency in making
essentially non-perturbative calculations is rather limited.

The supersymmetry approach makes use of the fact that the partition function
of non-interacting fermions is always the inverse of that of the analogous
bosonic theory. Therefore, if one introduces additional bosonic degrees of
freedom that replicate the fermionic action, the overall partition function
of the supersymmetric theory will be reduced to one. The approach is proven
to be a handy tool for studies in various fields of physics and in
particular, in models of quantum chaos involving random matrix theory and
various models of disorder \cite{Efetov-1983,Efetov-1997,Verbaarschot-1985}.

One of the prominent methods employing supersymmetry is the nonlinear
supersymmetric sigma model \cite{Efetov-1983, Efetov-1997} description of
disordered metallic conductors. According to this standard formalism,
effective field theory is described by action with coordinate dependent
supermatrix field, $Q(r)$, obeying the constraint, $Q^{2}(r)=1$. This method
has a broad range of applications, including the study of Anderson
localization, mesoscopic fluctuations, levels statistics in a limited
volume, quantum chaos. The limitation of the supersymmetric approach was
that it was deemed to be inapplicable to systems of interacting particles.

However, it turns out that there are important non-trivial models of
interacting particles with the disorder that can be written in a
supersymmetric form, and one can average over the disorder at the beginning
of calculations. The main goal of this paper is to identify such models and
develop the supersymmetry approach to the disorder averaging. To be more
specific, we will apply this approach to study the Sachdev-Ye-Kitaev (SYK)
model~\cite{Sachdev-1993, Kitaev-2015},originally considered in \cite{French, Bohigas}. 
In this model, the disorder averaging was so
far performed only within the replica trick approach. Our mapping of the SYK
model onto a supersymmetric model containing both fermion and boson degrees
of freedom subsequent averaging over the disorder is exact. Moreover, we
demonstrate that the new supersymmetric model with an effective
particle-particle interaction can be reformulated in terms of some
generalized supermatrix $\sigma $-model (superbosonization). This procedure
is also exact. Leaving investigation of new non-trivial regimes of the SYK
model for the future, we concentrate here on analyzing the semiclassical
limit of the model. The results obtained in the semiclassical limit within
this new approach are in agreement with those found earlier using the
replica technique. The applicability of the supersymmetry method to
the SYK model opens a new way of calculations for a certain class of models in condensed matter, gravity, and high energy physics.

The SYK model exhibits inherently non-Fermi liquid behavior and quantum
many-body chaotic eigenspectrum\cite%
{Sachdev-2015,Maldacena-2016,Jensen-2016,altland-2019prl,abk-2019,lunkin,bulucheva}
. This suggests that the two-point correlation function of the original
fields of the model does not fully capture the many-body level statistics.
The reason is that these are the many-body states that entirely determine
the close energy levels. Thus, the many-body level statistics of the SYK
model that follows the universal behavior of Wigner-Dyson random matrix
ensembles is inaccessible to original single-particle fields. To account for
many-body effects of the model, we perform the superbosonization
transformation and rewrite the model in terms of the collective many-body
excitations. To show the workability of the representation, we reproduce
earlier established results. We also demonstrate that the developed
superbosonized description of the SYK model is capable of producing novel
non-perturbative many-body effects.

The paper is organized as follows. In Section II we introduce the SYK model.
In Section III we develop a new, supersymmetric sigma-model representation
for interacting disordered fermion systems and apply it to SYK model. To
derive it, we decouple the interaction Hamiltonian using the conventional
Hubbard-Stratonovich approach. Then we notice that the Hubbard-Stratonovich
field can, in some situations, be gauged out from the denominator. This
enables one to supersymmetrize the interacting theory. In Section IV, the
new formalism is tested by calculating the fermion Green's function in the
SYK model at large times and is argued to be efficient for other interacting
models with the disorder.

In Section V we rewrite the supersymmetric SYK model as a model describing
unconstrained supermatrices representing collective many-body excitations.
Such a representation, where the partition function is represented in a
supermatrix action formulation without any constraints is dubbed
superbosonization. Since the transformation is exact, it is fully capable
describing the many-body modes instead of the original fermions of the SYK
model. As such, it represents the first step towards derivation of the
Wigner-Dyson eigenvalue statistics and the calculation of the Thouless time
at which the universal random-matrix behavior sets in. Our conclusions and
the possible directions for the future research are discussed in Section VI.


\section{Model}


The study of out-of-time correlation functions \cite{Larkin-1969} in the SYK
model shows \cite%
{Aleiner-2016,Stephen-Maldacena-2016,Maldacena-3-2016,Kitaev-2017,Bagrets-2017}%
, that it exhibits chaotic behavior at all time scales. At short times it
has exponentially decaying correlators, while at ultra-long times, otherwise
nearly zero-temperatures, when the energy scale is less than the many-body
level spacing, one has maximal chaoticity in the large system size limit.
This happens because Lyapunov exponent saturates to the conjectured upper
bound \cite{Stephen-Maldacena-2016}. One of the important problems here is
the test of Eigenstate Thermalization Hypothesis (ETH) \cite{Srednicki-1994,
Deutsch-1991}, which is a conjecture about the nature of matrix elements of
physical observables that, if holds, reconciles the predictions of
statistical physics of equilibrating states with those of quantum mechanics
in the longtime limit.

The study of low energy (long time) scale \cite{Altland-2017,Haque-1711}
shows ETH behavior because the eigenstates exhibit volume law entanglement
\cite{Liu-1709,yichen} suggesting that system becomes ergodic. However, one
of the major questions here is associated with finding the intermediate
time/energy scale, at which the system transfers to a thermalized state. The
characteristic time scale that leads to ergodicity in the SYK system is
analogous to Thouless time in dirty metals, while the states are analogous
to diffusive modes there. In the intermediate stage, one does not have an
ergodic state. To study the latter, in Refs.~\onlinecite
{Garcia-Garcia-2018,Danshita-2016,Haque-1711,You-2017}, the local two
fermions hopping term $(SYK_{2})$ with random coupling was added to the
four-fermion long-range randomly interacting $SYK_{4}$ Hamiltonian. Here the
thermalization properties, including the Lyapunov exponent (or scrambling
rate ) and the so-called butterfly velocity, were analyzed. The butterfly
speed is the speed at which the impacts of a local perturbation proliferate,
while the scrambling rate is a proportion of the rate at which the local
perturbation is mixed into non-local degrees of freedom. It has been
demonstrated that in a general quantum framework, the Lyapunov exponent is
limited by the temperature.

Another development in this direction was reported in Refs.~\onlinecite
{
khveshchenko-1705,khveshchenko-1805,Berkooz-2017,Gu-1609,Banerjee-2017,Jian-2017,Haldar-1703,Banerjee-1710,Jian-2-2017,Gu-2017,Song-2017,Chen-2017,Zhang-2017,Cai-2018,Zhong-1803,Mondal-1801,Dai-1802%
}, where d-dimensional generalization of SYK model was proposed by taking a
number of SYK droplets in real space and including fermion hopping terms
between them. This line of investigations is however out of the scope of the
present project.

The level statistics in the generalized SYK$_{4}$ + SYK$_{2}$ model was
studied recently using exact diagonalization \cite{Garcia-Garcia-2018}. The
results suggest that upon fixing the range of two-fermion hopping and
keeping the four-fermion interaction sufficiently long-ranged, the spectral
correlations will not change substantially compared to the random matrix
prediction, which is typical for chaotic quantum systems. However, by
reducing the range of the two-fermion terms, one will see a transition into
an insulating state, characterized by Poisson statistics. It appeared, that
in the vicinity of the many-body metal-insulator transition point, the
spectral correlations share all the features that had been previously found
in systems at the Anderson transition and in the proximity of the many-body
localization transition. This indicates the potential relevance of
generalized SYK models in the context of many-body localization and, also
exhibits itself as a starting point for the exploration of a gravity-dual of
this phenomenon.

An important demonstration of the SYK model being maximally chaotic is the
fact of having a finite entanglement entropy at zero temperature~\cite%
{Altland-2017,yichen}, indicating that at large time scales there is maximal
mixing in the ground state. Basic features of the SYK model, that in turn
support the existence of a gravity dual, include maximal chaos in the strong
coupling limit, finite zero temperature entropy, linear specific heat in the
low-temperature limit, the exponential growth of low-energy excitations, and
the short-range spectral correlations given by random matrix theory.

In its simplified version \cite{Sachdev-2015} the complex SYK model is a
system of randomly interacting $N$ (originally Majorana) spinless fermions
represented by their annihilation (creation) operators $\hat{c}_{i}$ ($\hat{c%
}_{i}^{\dagger }$), $i=1,\dots ,N$, with random all-to-all interactions
given by the Hamiltonian
\begin{equation}
\hat{H}=\sum_{ij,kl}J_{ij,kl}\hat{c}_{i}^{\dagger }\hat{c}_{j}\hat{c}%
_{k}^{\dagger }\hat{c}_{l}-\mu \sum_{i}\hat{c}_{i}^{\dagger }\hat{c}_{i}.
\label{e1}
\end{equation}%
The coupling constant $J_{ij,kl}$ were assumed to be random complex number
\begin{equation}
\;J_{ij,kl}^{\ast }=J_{lk,ji}  \label{e1a}
\end{equation}%
with a Gaussian distribution characterized by the following average and
variance
\begin{equation}
\left\langle J_{ij,kl}\right\rangle =0,\;\langle J_{ij,kl}J_{i^{\prime
}j^{\prime },k^{\prime }l^{\prime }}^{\ast }\rangle =\frac{J^{2}}{N^{3}}%
\delta _{ii^{\prime }}\delta _{jj^{\prime }}\delta _{kk^{\prime }}\delta
_{ll^{\prime }}.  \label{e2}
\end{equation}
Averages of the type $\langle J_{ij,kl}J_{i^{\prime }j^{\prime },k^{\prime
}l^{\prime }}\rangle $ are equal to zero unless they can be reduced to Eq. (%
\ref{e2}) using the symmetry relation (\ref{e1a}).
The generalization of the SYK model to the case of random $q$-fermion interaction with even $q$ is dubbed in the literature $SYK_q$ model. In the latter,  instead of four fermion interactions with random coupling, one has $q$ fermion interaction.

At large time scales (low temperatures) the SYK model is conformal because
the term that contains a time-derivative in the Lagrangian can be ignored.
The action of the model can be written using the so-called $G,\Sigma $
representation, and the Schwarzian theory \cite{Kitaev-2015, Maldacena-2016}
can describe its soft mode fluctuations. It has been shown that this theory
is equivalent \cite{Maldacena-2-2016,Jensen-2016, Engelsoy-2016} to
two-dimensional dilaton gravity, the Jackiw-Teitelboim model \cite%
{Jakiw-1985,Teitelbom-1983,Almheiri-2015}. This fact points out the link
between AdS$_{2}$ black hole physics and the SYK model.

The spectral form factor in the SYK model was studied numerically in Ref.~%
\onlinecite{Cotler-2017} (analogous two-point correlation functions were
studied using the random matrix approach in Refs~\onlinecite
{You-2017,Garsia-Garsia-2016}). Roughly, the spectral form factor is the
Fourier transform of the connected two-point density-density correlation
function $\langle \rho (E)\rho (E^{\prime })\rangle $ in random matrix
theory. The question of the precise window of universality in which random
matrix theory is applicable is still unknown.

The supersymmetric reformulation of the SYK model we propose below, may give
a possibility to derive it theoretically. The method may also allow one to
study corrections beyond this universality regime. It is worth emphasizing
that developing the supersymmetric representation we start with a fermionic
SYK model analogous to the one given by Eq.~(\ref{e1}). Bosons appear after
certain transformations and are somehow \textquotedblleft fictitious
bosons\textquotedblright\ like those that appear in the supersymmetry
technique for electron systems \cite{Efetov-1983,Efetov-1997}.

All this clearly contrasts works on supersymmetric generalizations of the
SYK model. For example, Ref.~ \onlinecite{Fu-2017} reported a supersymmetric
generalization of the SYK. In that model, however, the four fermion coupling
constants $J_{ijkl}$ are not entirely random (they are correlated and
defined by free coupling constants in supercharge $Q$). The bosonic field
appears here as a non-dynamical field to linearize the supersymmetry
transformation and realize the supersymmetry algebra off-shell. Similar
supersymmetric lattice models were reported in Refs.
\onlinecite{Fendley-2003, Fendley-2003-2, Fendley-2005,
Huijse-2008,Huijse-2010, Huijse-2011, Huijse-2012}. Specific correlations of
the random couplings of these models lead to $\mathcal{N}=1$ and $\mathcal{N}%
=2$ supersymmetry. Previous attempts to use a supersymmetry technique for the 
$k$-body matrix models were reported in \cite{Weidenmuller}. 
Supersymmetric models with random couplings that include
both bosons and fermions were considered in Refs.~%
\onlinecite{Anninos-2016,Murugan1}, while Ref.~\onlinecite{Murugan2}
explored the possibility of extending the 1+1 dimensional bosonization
technique to (0 + 1)-dimensional SYK-type systems. Ref.~%
\onlinecite{1908.00995} suggested that the SYK model with Majorana fermions
and without fine-tuned couplings has the capacity of possessing some hidden
supersymmetry, which may also be present in complex SYK model when the
chiral symmetry is present\cite{1912.09975}.


\section{Supersymmetry reformulation of SYK model: averaging over quenched
disorder}


Now we apply the supersymmetry approach to the interacting SYK model. We
believe that such an approach opens the door to analyzing the many-body
effects and exponentially small bulk level spacing of the model. The
formalism could also be adapted to study the effects in generalized SYK
models such as SYK$_{4}$ + SYK$_{2}$ and establish a fruitful connection
between complex and Majorana models.

Although the original model, Eq. (\ref{e1}), has been written in the
Hamiltonian representation, it is more convenient to use the functional
integral representation with fermionic fields $\chi _{i}\left( t\right) $, $%
\chi _{i}^{\ast }\left( t\right).$ They obey the anticommutation relations
\begin{equation}
\left\{ \chi _{i},\chi _{j}\right\} =\left\{ \chi _{i}^{\ast },\chi
_{j}^{\ast }\right\} =\left\{ \chi _{i},\chi _{j}^{\ast }\right\} =0,
\label{e2a}
\end{equation}
and we use the convention $\left( \chi _{i}^{\ast }\right) ^{\ast }=-\chi
_{i}$.

In order to develop the supersymmtery approach for the model with the
fermion-fermion interaction, we slightly modify the original model specified
by Eqs.~(\ref{e1}). Using the anticommuting Grassmann fields, $\chi $, we
write correlation functions in terms of a functional integral over these
fields as
\begin{equation}
G_{ij}\left( t,t^{\prime }\right) =-\frac{i\int \chi_i \left( t\right)
\chi_j ^{\ast }\left( t^{\prime }\right) \exp \left( iS\left[ \chi ,\chi
^{\ast }\right] \right) D\chi D\chi ^{\ast }}{\int \exp \left( iS\left[ \chi
,\chi ^{\ast }\right] \right) D\chi D\chi ^{\ast }}  \label{e3}
\end{equation}
In Eq. (\ref{e3}), the product of the fields $\chi_i \left( t\right) $ and $%
\chi_j ^{\ast }\left( t\right) $ for arbitrary $i,j$ and times $t$ defines
the Green's function $G_{ij}$. Here we start with the action, $S\left[ \chi
,\chi ^{\ast }\right] $, which is slightly different from the field
representation of the model given by Eq. (\ref{e1}). Namely, we consider
\begin{eqnarray}
&&S\left[ \chi ,\chi ^{\ast }\right] =\int_{-\infty }^{\infty }\big[ %
\sum_{i=1}^{N}\chi _{i}^{\ast }\left( i\partial _{t}+\mu \right) \chi
_{i}\left( t\right)  \label{e4} \\
&&-\sum_{ij,kl=1}^{N}J_{ij,kl}\left( \chi _{i}^{\ast }\left( t\right) \chi
_{j}\left( t\right) -\chi _{j}^{\ast }\left( t\right) \chi _{i}\left(
t\right) \right)  \notag \\
&&\times \left( \chi _{k}^{\ast }\left( t\right) \chi _{l}\left( t\right)
-\chi _{l}^{\ast }\left( t\right) \chi _{k}\left( t\right) \right) \big]dt.
\notag
\end{eqnarray}
The random coupling constants $J_{ij,kl}$ in Eq. (\ref{e4}) are assumed to
be real and obey the symmetry relations
\begin{equation}
J_{ij,kl}=-J_{ji,kl}=-J_{ij,lk}=J_{kl,ij}.  \label{e4b}
\end{equation}
Their distribution is Gaussian with zero average
\begin{equation}
\left\langle J_{ij,kl}\right\rangle =0,  \label{e4c}
\end{equation}
and the variance
\begin{eqnarray}
&&\left\langle J_{ij,kl}J_{i^{\prime }j^{\prime },k^{\prime }l^{\prime
}}\right\rangle =\frac{J^{2}}{8N^{3}}  \notag \\
&&\times \big(\left( \delta _{ii^{\prime }}\delta _{j,j^{\prime }}-\delta
_{ij^{\prime }}\delta _{ji^{\prime }}\right) \left( \delta _{k,k^{\prime
}}\delta _{ll^{\prime }}-\delta _{kl^{\prime }}\delta _{lk^{\prime }}\right)
\notag \\
&&+\left( \delta _{ik^{\prime }}\delta _{j,l^{\prime }}-\delta _{il^{\prime
}}\delta _{jk^{\prime }}\right) \left( \delta _{k,i^{\prime }}\delta
_{lj^{\prime }}-\delta _{kj^{\prime }}\delta _{li^{\prime }}\right) \big).
\label{e4d}
\end{eqnarray}
One can interpret the model described by Eq. (\ref{e4}) as a time-reversal
invariant symmetrized version of the SYK model. The model was also recently considered in Ref.~\onlinecite{Debanjan}.

First, under the functional integral, we introduce a time-dependent
Hubbard-Stratonovich real antisymmetric matrix field, $M^F_{ij}\left(
t\right) $, and decouple the four-fermion interaction of the SYK Hamiltonian
(\ref{e1a}) by inserting identity operator,
\begin{eqnarray}
&&\mathds{1}\equiv \int \frac{\mathcal{D}M^{F}}{Det[J_{ij,kl}]}\exp \Big\{%
i\int dt  \label{HS} \\
&&\sum_{ij,kl=1}^{N}\sum_{i^{\prime }j^{\prime },k^{\prime }l^{\prime
}=1}^{N}\big(M_{ij}^{F}-i\left( \chi _{l}^{\ast }\chi _{k}-\chi _{k}^{\ast
}\chi _{l}\right) J_{kl,ij}\big)  \notag \\
&&\times (J^{-1})_{ij,i^{\prime }j^{\prime }}\big(M_{j^{\prime }i^{\prime
}}^{F}-J_{i^{\prime }j^{\prime },k^{\prime }l^{\prime }}i\left( \chi
_{k^{\prime }}^{\ast }\chi _{l^{\prime }}-\chi _{l^{\prime }}^{\ast }\chi
_{k^{\prime }}\right) \big)\Big\},  \notag
\end{eqnarray}%
into the functional integrals over $\chi ,\chi ^{\ast }$ in Eq. (\ref{e3}).
Here $(J^{-1})_{ij,kl}$ is the inverse of $J_{ij,kl}$, namely
\begin{equation}
\sum_{kl}(J^{-1})_{ij,kl}J_{kl,mn}\mathbf{=\delta }_{im}\mathbf{\delta }%
_{jn}.  \label{e4a}
\end{equation}
Using the last property in Eqs. (\ref{e4b}) of the coupling $J_{ij,kl}$ and
the hermiticity of the matrix, $M_{ij}\left( t\right),$ we see that the
exponent in Eq.~(\ref{HS}) is purely imaginary and the integral over matrix $%
M_{ij}\left( t\right)$ converges. Then the action for the time-reversal
symmetric modification of the SYK model is now equivalent to that of a
system of electrons moving in a fluctuating real antisymmetric field $%
M_{ij}\left( t\right) $ with random $J_{ij,kl}$:
\begin{eqnarray}
&&S\left[ \chi ,\chi ^{\ast },M^{F}\right] = S_0\left[ \chi ,\chi ^{\ast },M^{F}\right]+S_{fluct}\left[M^{F}\right]\notag \\
&=&\int_{-\infty }^{\infty }dt\sum_{ij=1}^{N}\Big\{\chi _{i}^{\ast }\left(
t\right) \big[(i\partial _{t}+\mu )\delta _{ij}-2iM_{ij}^{F}\left( t\right) %
\big]\chi _{j}\left( t\right)  \notag \\
&&+\sum_{ijkl}M_{ij}^{F}\left( t\right) (J^{-1})_{ij,kl}M_{lk}^{F}\left(
t\right) \Big\}.  \label{e5}
\end{eqnarray}
Here $S_{fluct}\left[M^{F}\right]\notag$ represents the Gaussian fluctuations of $M_F(t)$. This action therefore defines the Green's function, $G_{ij}$, of fermionic
fields, $\chi_i\left( t\right) ,\chi_j ^{\ast }\left( t\right)$, as
\begin{eqnarray}  \label{e6}
&&G_{ij}\left( t,t^{\prime }\right) = \\
&&-\frac{i\int \chi _{i}\left( t\right) \chi _{j}^*\left( t^{\prime }\right)
\exp \left( iS\left[ \chi ,\chi ^{\ast},M^{F}\right] \right) D\chi D\chi
^{\ast }DM^{F}}{\int \exp \left( iS\left[ \chi ,\chi ^{\ast },M\right]
\right) D\chi D\chi ^{\ast}DM}.  \notag
\end{eqnarray}

The random coupling $J_{ij,kl}$ enters both the numerator and denominator in
Eq.~(\ref{e6}), and one cannot average over this coupling directly. This
situation is typical for problems with quenched disorder. The standard
supersymmetry approach of Refs. \onlinecite{Efetov-1983,Efetov-1997} relies
on the fact that the system is initially non-interacting. In that case, one
replaces the denominator by an integral over bosonic fields in the
numerator. Since here we deal with an inherently interacting system, we
generate a field $M_{ij}$ which enters both numerator and denominator in Eq.
(\ref{e6}) and seemingly invalidates the possibility of supersymmetrizing
the action.

Although this obstacle cannot be generally overcome, the SYK model
considered here is in this respect exceptional. Now we make a crucial
observation. We show now that the integral over the fermionic fields $\chi
,\chi ^{\ast }$ in denominator of Eq.~(\ref{e6}) does not, in fact, depend
on the Hubbard-Stratonovich field $M(t)$. 
The reason is that the real antisymmetric matrix $M(t)$ 
can be reduced to a time-independent constant matrix $M_0$
by a gauge transformation $2M(t)=2 U^{T}M_0 U -U^{T}\partial _{t}U$ of the orthogonal group $U^{T}U=1$. Here the constant matrix $M_0$ is block diagonal with real $2 \times 2$ antisymmetric blocks along the diagonal 
$\hat{\mu_i} = \Big(\begin{array}{cc}
0 & \mu_i \\
-\mu_i & 0
\end{array}
\Big)$, with $i=1,2,\cdots N/2$. The matrix $M_0$ represents the zero-mode of $M(t)$ and appears due to periodicity of restrictions on $U(t)$.  At zero temperature it vanishes, $M_0=0$, and the transformation reduces to  a pure gauge transformation $2 M(t)\rightarrow -U^{T}\partial _{t}U$. 
Since the gauge transformation of free fermions is not anomalous [\onlinecite{Elitzur-1986}], it helps us to simplify the integral
in the denominator of Eq.~(\ref{e6}):
\begin{eqnarray} \label{denominator}
&&\int \exp \left( iS_0\left[ \chi ,\chi ^{\ast },M\right] \right) D\chi D\chi
^{\ast } \label{PF} \\
&=&Det[(i\partial _{t}+\mu )\delta _{ij}-2iM_{ij}] \notag \\
&=&Det[U^{T}(i\partial _{t}+\mu )U]=Det[(i\partial _{t}+\mu)]. \notag
\end{eqnarray} 
So, what we end up having in the denominator is just a determinant, 
$Det[(i\partial _{t}+\mu)]$, which is independent of the
fluctuating field $M(t)$ and random coupling constants $J_{ijkl}$.
 This point is crucial and it
allows one to express the integral over the fermionic fields in the
denominator in Eq. (\ref{e6}) via additional bosonic \textit{superpartner}
fields. This is a standard procedure of the supersymmetric approach
developed in Refs.~\onlinecite
{Efetov-1983,Efetov-1997}.

Following this approach, we introduce complex bosonic fields $s_{i}\left(
t\right) $, $i=1,2...N$ and a new bosonic model with the action
\begin{eqnarray}
&&S^{B}\left[ s,s^{\ast }\right] =\int_{-\infty }^{\infty }\big[%
\sum_{i=1}^{N}s_{i}^{\ast }\left( t\right) \left( i\partial _{t}+\mu \right)
s_{i}\left( t\right)  \label{k1} \\
&&-\sum_{ij,kl=1}^{N}J_{ij,kl}\left( s_{i}^{\ast }\left( t\right)
s_{j}\left( t\right) -s_{j}^{\ast }\left( t\right) s_{i}\left( t\right)
\right)  \notag \\
&&\times \left( s_{k}^{\ast }\left( t\right) s_{l}\left( t\right)
-s_{l}^{\ast }\left( t\right) s_{k}\left( t\right) \right) \big]dt.  \notag
\end{eqnarray}
The action $S^{B}\left[ s,s^{\ast }\right] $ looks identical to action $S%
\left[ \chi ,\chi ^{\ast }\right] $, Eq.~(\ref{e4}), and it is real.
Moreover, the coupling constant $J_{ij,kl}$ obeys the same symmetry
relations (\ref{e4b}). Now we write the bosonic partition function
\begin{equation}
Z_{B}=\int \exp \left( iS\left[ s,s^{\ast }\right] \right) DsDs^{\ast }.
\label{k2}
\end{equation}
As the action $S^{B}\left[ s,s^{\ast }\right] $ is real, the integral over $%
s\left( t\right) $ in Eq.~(\ref{k2}) converges. Then, we make the same
decoupling of the interaction in Eq. (\ref{k2}) as we have done for the
fermionic model, and write the partition function $Z_{B}$ in the form
\begin{equation}
Z_{B}=\int \exp \left( iS^{B}\left[ s,s^{\ast },M^{B}\right] \right)
DsDs^{\ast }DM^{B}.  \label{k3}
\end{equation}%
Here the action $S^{B}\left[ s,s^{\ast },M^B\right] $ equals to
\begin{eqnarray}
&&S\left[ s,s^{\ast },M^{B}\right]  \notag \\
&=&\int_{-\infty }^{\infty }dt\sum_{i,j=1}^{N}\Big\{s_{i}^{\ast }\left(
t\right) \big[(i\partial _{t}+\mu )\delta _{ij}-2iM_{ij}^{B}\left( t\right) %
\big]s_{j}\left( t\right)  \notag \\
&&+\sum_{ij,kl}M_{ij}^{B}\left( t\right) (J^{-1})_{ij,kl}M_{kl}^{B}\left(
t\right) \Big\}.  \label{k4}
\end{eqnarray}%
The matrix $M^{B}\left( t\right) $ in Eqs.~(\ref{k3}) and (\ref{k4}) has the
same symmetry as the matrix $M^{F}\left( t\right)$ in Eqs.~(\ref{HS})
through (\ref{PF}), and we can calculate the Gaussian integrals over the
bosonic field $\mathbf{s}\left( t\right) $ in the same manner as previously:
\begin{eqnarray}
&&\int \exp \Big[i\int_{-\infty }^{\infty }\big[s_{i}^{\ast }\left( t\right)
(i\partial _{t}+\mu )\delta _{ij}  \label{e8a} \\
&&-2M_{ij}^{B}\left( t\right) \big]s_{j}\left( t\right) dt\Big]DsDs^{\ast }
\notag \\
&=&\left( Det[(i\partial _{t}+\mu )\delta _{ij}-2M_{ij}^{B}]\right)
^{-1}=\left( Det(i\partial _{t}+\mu )\right) ^{-1}.  \notag
\end{eqnarray}%
We see that the matrix $M^{B}\left( t\right) $ is gauged out, and the result
of the integration over $s\left( t\right) ,$ $s^{\ast }\left(t\right) $ is
performed exactly in the same way as in the fermionic determinant. This
matrix is also real and antisymmetric. However, in contrast to Eq.~(\ref{PF}%
), one obtains the inverse of the determinant. It is this property of
bosonic determinants that allows one to get rid of the denominator in Eq.~(%
\ref{e3}).

Combining the fermionic and bosonic degrees of freedom, one can form a
supervector $\Phi \equiv (\{\chi _{i}\};\left\{ s_{i}\right\} )\in
U(N,1|N,1)$ and its Hermitian conjugate supervector $\Phi ^{\dagger}\in U(N,1|N,1)$%
. This allows us to write a supersymmetric action for the time-reversal
invariant SYK model as
\begin{eqnarray}
&&\tilde{S}\left[ \Phi ,\Phi ^{\dagger},\hat{M}\right]  \notag \\
&=&\int dt\Big[\sum_{i,a}\Phi _{i}^{\dagger}\left( t\right) \big[(i\partial
_{t}+\mu )\delta _{ij}-2\hat{M}_{ij}\left( t\right) \big]\Phi _{j}\left(
t\right)  \notag \\
&+&\sum_{ijkl}\mathrm{Tr}\left( \hat{M}_{ij}\left( t\right) (J^{-1})_{ij,kl}%
\hat{M}_{kl}\left( t\right) \right) \Big],  \label{e9}
\end{eqnarray}%
where the two-component supervectors have the following structure
\begin{equation}
\Phi _{i}\left( t\right) =\left(
\begin{array}{c}
\chi _{i}\left( t\right) \\
s_{i}\left( t\right)%
\end{array}%
\right) ,\;\Phi _{i}^{\dagger}\left( t\right) =\left(
\begin{array}{cc}
\chi _{i}^{\ast }\left( t\right) & s_{i}^{\ast }\left( t\right)%
\end{array}%
\right) .  \label{e10}
\end{equation}%
and
\begin{equation}
\hat{M}_{ij}\left( t\right) =\left(
\begin{array}{cc}
M_{ij}^{F}\left( t\right) & 0 \\
0 & M_{ij}^{B}\left( t\right)%
\end{array}%
\right)  \label{e11}
\end{equation}%
is a diagonal matrix in the space of the supervectors, $\Phi $. Having set
up this structure, one can readily write the fermion Green's function $%
G_{ij} $, in Eq. (\ref{e3}), as
\begin{eqnarray}
&&G_{ij}\left( t,t^{\prime }\right) =-i\int \Phi _{i}^{1}\left( t\right)
\Phi _{j}^{1\dagger}\left( t^{\prime }\right)  \label{e12} \\
&&\times \exp \left( iS\left[ \Phi ,\Phi ^{\dagger},\hat{M}\right] \right) D\Phi
D\Phi ^{\dagger}D\hat{M}.  \notag
\end{eqnarray}
Importantly, the absence of the weight denominator in Eq.~(\ref{e12}) allows
one to average over the random coupling $J_{ij,kl}$ in the beginning of all
calculations.

We see that, although we have started with an interacting theory, the
supersymmetry approach to quenched averaging\cite{Efetov-1997} works in this
case as well due to the fact that the spatial dimension in this problem is
effectively zero. We emphasize that all the transformations reducing Eq.~(%
\ref{e3}) to Eqs.~(\ref{e12}) are formally exact. Now we integrate in Eqs.~%
(\ref{e12}) over the matrix $\hat{M}\left(t\right) $ to obtain
\begin{equation}
G_{ij}\left( t,t^{\prime }\right) =\int \Phi _{i}^{1}\left( t\right) \Phi
_{j}^{1\dagger}\left( t^{\prime }\right) \exp \left( iS\left[ \Phi ,\Phi ^{\dagger} %
\right] \right) D\Phi D\Phi ^{\dagger},  \label{e14}
\end{equation}
Here, in Eq.~(\ref{e14}), the action $S\left[ \Phi ,\Phi ^{\dagger}\right] $
equals
\begin{eqnarray}
&&S\left[ \Phi ,\Phi ^{\dagger}\right] =\int_{-\infty }^{\infty }dt\Big[%
\sum_{i}\Phi _{i}^{\dagger}\left( t\right) (i\partial _{t}+\mu )\Phi _{i}\left(
t\right)  \notag \\
&&-\sum_{a=1}^{2}\sum_{ij,kl}J_{ij,kl}\left( \Phi _{i}^{a\dagger}\left( t\right)
\Phi _{j}^{a}\left( t\right) -\Phi _{j}^{a\dagger}\left( t\right) \Phi
_{i}^{a}\left( t\right) \right)  \notag \\
&&\times \left( \Phi _{k}^{a\dagger}\left( t\right) \Phi _{l}^{a}\left( t\right)
-\Phi _{l}^{a\dagger}\left( t\right) \Phi _{k}^{a}\left( t\right) \right) \Big],
\label{e16}
\end{eqnarray}%
where $a=1$ denotes the fermionic component of the supervector $\Phi \left(
t\right) $ defined in Eq.~(\ref{e10}), while $a=2$ stands for the bosonic
one.

Before performing disorder averaging, it is convenient to use more compact
notations via introducing $4$- component supervectors $\Psi \left( t\right) $
as
\begin{eqnarray}
\Psi _{i}\left( t\right) &=&\frac{1}{\sqrt{2}}\left(
\begin{array}{c}
\chi _{i}^{\ast }\left( t\right) \\
\chi _{i}\left( t\right) \\
s_{i}^{\ast }\left( t\right) \\
s_{i}\left( t\right)%
\end{array}
\right) ,\;  \label{e16a} \\
\bar{\Psi}_{i} &=&\frac{1}{\sqrt{2}}\left(
\begin{array}{cccc}
\chi _{i}\left( t\right) & \chi _{i}^{\ast }\left( t\right) & -s_{i}\left(
t\right) & s_{i}^{\ast }\left( t\right)%
\end{array}%
\right) .  \notag
\end{eqnarray}
The supervector $\bar{\Psi}$ is related to $\Psi $ by a charge conjugation:
\begin{equation}
\bar{\Psi}=\left( C\Psi \right) ^{T},  \label{e16b}
\end{equation}
where \textquotedblleft $T$\textquotedblright\ stands for transposition, and
the matrix $C$ is given by
\begin{equation*}
C=\left(
\begin{array}{cc}
c_{2} & 0 \\
0 & c_{1}%
\end{array}%
\right) ,\;c_{1}=\left(
\begin{array}{cc}
0 & -1 \\
1 & 0%
\end{array}%
\right) ,\;c_{2}=\left(
\begin{array}{cc}
0 & 1 \\
1 & 0%
\end{array}%
\right) .
\end{equation*}%
One can notice that $\bar{\Psi}$ has a simple connection to the hermitian
conjugated supervector $\Psi ^{\dagger}$:
\begin{equation}
\bar{\Psi}=\Psi ^{\dagger}\tau _{3},  \label{e16c}
\end{equation}%
where
\begin{equation}
\tau _{3}=\left(
\begin{array}{cc}
1 & 0 \\
0 & -1%
\end{array}
\right)  \label{e16d}
\end{equation}
is Pauli matrix in the \textquotedblleft particle-hole\textquotedblright\
space of matrices $c_{2}$ and $c_{1}$.

Furthermore, the square of the modulus of the supervector $\Psi $ is equal
to
\begin{equation}
\left\vert \Psi \right\vert ^{2}=\Psi ^{\dagger}\Psi =\bar{\Psi}\tau _{3}\Psi .
\label{e16e}
\end{equation}
It is also seen that
\begin{equation}
\bar{\Psi}_{i}\Psi _{j}=-\bar{\Psi}_{j}\Psi _{i}.  \label{e16f}
\end{equation}
Substituting Eqs.~(\ref{e16a}) through (\ref{e16f}) into Eqs. (\ref{e14})
and (\ref{e16}), we rewrite the fermion Green's function in a more compact
form
\begin{equation}
G_{ij}\left( t,t^{\prime }\right) =\int \Psi _{2i}^{1}\left( t\right) \Psi
_{2j}^{1\dagger}\left( t^{\prime }\right) \exp \left( iS\left[ \Psi ,\Psi ^{\dagger}%
\right] \right) D\Psi D\Psi ^{\dagger}.  \label{e16g}
\end{equation}
In Eq. (\ref{e16g}), superscripts numerate blocks in the superspace, while
first subscripts numerate elements in the particle-hole space. The action $S%
\left[ \Psi ,\Psi ^{\dagger}\right] $ entering Eq.~( \ref{e16g}) is given by
\begin{eqnarray}
&&S\left[ \Psi ,\Psi ^{\dagger}\right] =\int_{-\infty }^{\infty }dt\Big[\sum_{i}%
\bar{\Psi}_{i}\left( t\right) (i\partial _{t}+\tau _{3}\mu )\Psi _{i}\left(
t\right) \qquad\;\;\;  \notag \\
&&-4\sum_{a=1}^{2}\sum_{ij,kl}J_{ij,kl}\left( \bar{\Psi}_{i}^{a}\left(
t\right) \Psi _{j}^{a}\left( t\right) \right) \left( \bar{\Psi}%
_{k}^{a}\left( t\right) \Psi _{l}^{a}\left( t\right) \right) \Big],
\label{e16h}
\end{eqnarray}%
where $\Psi ^{a}\left( t\right) $, $a=1,2$ stand for the fermion and boson
components of the supervectors $\Psi$. Substituting Eq.~(\ref{e16h}) into
Eq.~(\ref{e16g}), one can easily average over the random $J_{ij,kl}$ using
Eq. (\ref{e4d}). The expression for disorder averaged Green's function thus
will read as \textbf{\ }%
\begin{equation}
\left\langle G_{ij}\left( t,t^{\prime }\right) \right\rangle =\int \Psi
_{2i}^{1}\left( t\right) \Psi _{2j}^{1\dagger}\left( t^{\prime }\right) \exp
\left( i\bar{S}\left[ \Psi ,\Psi ^{\dagger}\right] \right) D\Psi D\Psi ^{\dagger},
\label{e17}
\end{equation}
where the non-local action action $\bar{S}\left[ \Psi ,\Psi ^{\dagger} \right] $
equals
\begin{eqnarray}
&&\bar{S}\left[ \Psi ,\Psi ^{\dagger}\right] =\int_{-\infty }^{\infty
}dt\sum_{i=1}^{N}\bar{\Psi}_{i}\left( t\right) (i\partial _{t}+\tau _{3}\mu
)\Psi _{i}\left( t\right)  \notag \\
&&\times \frac{iJ^{2}}{N^{3}}\sum_{a,b=1}^{2}\sum_{ij,kl}^{N}\int_{-\infty
}^{\infty }\left( \bar{\Psi}_{i}^{a}\left( t\right) \Psi _{j}^{a}\left(
t\right) \right) \left( \bar{\Psi}_{k}^{a}\left( t\right) \Psi
_{l}^{a}\left( t\right) \right)  \notag \\
&&\times \left( \bar{\Psi}_{l}^{b}\left( t^{\prime }\right) \Psi
_{k}^{b}\left( t^{\prime }\right) \right) \left( \bar{\Psi}_{j}^{b}\left(
t^{\prime }\right) \Psi _{i}^{b}\left( t^{\prime }\right) \right)
dtdt^{\prime }.  \label{S}
\end{eqnarray}
We see that the action ${\bar{S}}\left[ \Psi ,\Psi ^{\dagger}\right] $ in Eq. (\ref%
{S}) does not contain disorder anymore, and the integral over the
supervectos $\Psi _{i}^{\dagger}\left( t\right) $ and $\Psi _{j}\left( t\right) $
in Eq. (\ref{e17}) is clearly convergent.

Of course, in Eq.~(\ref{S}) the addition of extra bosonic degrees of freedom
comes at the price of introducing additional integrals. However, the
resultant theory, Eqs.~(\ref{e17}) and (\ref{S}), does not contain disorder
and is fully supersymmetric. As such, it has many simplifications. One
simplification is the cancellation of a variety of Feynman diagrams in the
perturbation theory in interactions due to the supersymmetry. Another
simplification follows from the superbosonization of this supersymmetric
action discussed in Section~V. In the superbosonized representation, instead
of the functional integral over supervectors, one deals with an integral
over supermatrices. In that approach, the number of integration variables
can significantly be reduced upon the diagonalization of the supermatrices.

However, let us fist make a saddle point approximation that has to become
exact in the limit $N\rightarrow \infty $. This is done in the next section.
Comparison of the hereby obtained results with those obtained within the
replica approach in Refs.~\onlinecite{Sachdev-1993} and %
\onlinecite{Sachdev-2015} can be done but one cannot expect a full
coincidence because we consider a somewhat different model. In contrast to
the calculations presented there we use the real-time representation.


\section{Saddle-point approximation}


The saddle point approximation is expected to become exact in the limit $%
N\rightarrow \infty $. In order to see this property explicitly and proceed
with the calculations, let us introduce $2\times 2$ supermatrices, $%
W^{ab}\left( t,t^{\prime }\right) $ as
\begin{equation}
W^{ab}\left( t,t^{\prime }\right) =\frac{2}{N}\sum_{i=1}^{N}\Psi
_{i}^{a}\left( t\right) \bar{\Psi}_{i}^{b}\left( t^{\prime }\right) ,
\label{e18}
\end{equation}
where supervectors $\Psi $ and $\bar{\Psi}$ are specified in Eq.~(\ref{e10}).
The supermatrix $W\left( t,t^{\prime }\right) $ has the evident symmetry
\begin{equation}
W^{\dagger}\left( t,t^{\prime }\right) =W\left( t^{\prime },t\right).  \label{e19}
\end{equation}
Using Eqs.~(\ref{e18}) and (\ref{e19}), and the disorder averaging procedure
resulting in Eq. (\ref{S}), we explicitly reduce Eq. (\ref{S}) to a
considerably more compact form
\begin{eqnarray}
&&\bar{S}\left[ \Psi ,\Psi ^{\dagger}\right] =\int_{-\infty }^{\infty
}dt\sum_{i=1}^{N}\bar{\Psi}_{i}\left( t\right) (i\partial _{t}+\tau _{3}\mu
)\Psi _{i}\left( t\right)  \notag \\
&&+\frac{iNJ^{2}}{2}\int_{-\infty }^{\infty }\int_{-\infty }^{\infty
}dtdt^{\prime }\Big[2\left( \mathrm{Tr}\left( W^{21}\left( t,t^{\prime
}\right) W^{12}\left( t^{\prime },t\right) \right) \right) ^{2}  \notag \\
&&+\left( \mathrm{Tr}\left( W^{11}\left( t,t^{\prime }\right) W^{11}\left(
t^{\prime },t\right) \right) \right) ^{2}  \notag \\
&&+\left( \mathrm{Tr}\left( W^{22}\left( t,t^{\prime }\right) W^{22}\left(
t^{\prime },t\right) \right) \right) ^{2}\Big].  \label{k6}
\end{eqnarray}%
where $2\times 2$ matrices ${W}^{ab}$ have matrix-valued entries. Elements
of matrices $W^{21}\left( t,t^{\prime }\right)$ and $W^{12}\left( t^{\prime
},t\right) $ are anticommuting fields, while those of the matrices $%
W^{11}\left( t,t^{\prime }\right)$ and $W^{22}\left( t,t^{\prime }\right)$
contain products of two anticommuting fields or are conventional complex
functions.

Here we would like to invite the reader's attention to the resemblance of
the action (\ref{S}) with the replicated imaginary time action of the SYK
model outlined in Ref.~\onlinecite{Sachdev-2015} (see Eq. (16) there).
However, now we have the formally exact supersymmetric representation of the
model, where no replica limit, $n\rightarrow 0$ (see e.g., Refs.~%
\onlinecite
{km, kanzieper, Sedrakyan-2005}), has to be taken. It is also worth
emphasizing that here we have $4\times 4$ supermatrices $W\left( t,t^{\prime
}\right) $ instead of $n\times n$ matrices in the replica approach. We
emphasize that Eqs. (\ref{e17}) and (\ref{k6}) are still exact for any $N$.

Now one can explicitly see that the interaction term in Eq.~(\ref{k6}) is
proportional to $N$, and the accuracy of the saddle-point approximation
should follow from the assumption that this number is large. Although
details are different, we use the general chain of transformations suggested
in Refs.~\onlinecite{Sachdev-1993} and \onlinecite{Sachdev-2015} and analyze
the behavior of the fermion Green's function.

First, we decouple the interaction terms in Eq.~(\ref{k6}) by introducing
auxiliary functions $P^{ab}\left( t,t^{\prime }\right)$ and integrating over
them. We write
\begin{eqnarray}
&&\exp \Big[-J^{2}N\sum_{a,b=1}^{2}\int_{-\infty }^{\infty }\int_{-\infty
}^{\infty }  \notag \\
&&\times \frac{1}{2}\left[ \mathrm{Tr}\left( W^{ab}\left( t,t^{\prime
}\right) W^{ba}\left( t^{\prime },t\right) \right) \right] ^{2}dtdt^{\prime }%
\Big]  \notag \\
&=&Z_{0}\int DP\exp \Big[-N\int_{-\infty }^{\infty }\int_{-\infty }^{\infty
}dtdt^{\prime }  \notag \\
&&\times \sum_{a,b=1}^{2}\Big[\frac{P^{ab}\left( t,t^{\prime }\right)
P^{ba}\left( t^{\prime },t\right) }{2J^{2}}  \notag \\
&&-iP^{ab}\left( t,t^{\prime }\right) \mathrm{Tr}\left( W^{ab}\left(
t^{\prime },t\right) W^{ba}\left( t,t^{\prime }\right) \right) \Big]\Big],
\label{e21}
\end{eqnarray}%
where
\begin{equation}
Z_{0}=\int DP\exp \Big[-N\sum_{a,b=1}^{N}\int_{-\infty }^{\infty }\frac{%
P^{ab}\left( t,t^{\prime }\right) P^{ba}\left( t^{\prime },t\right) }{2J^{2}}%
dtdt^{\prime }\Big].  \label{e21a}
\end{equation}%
In Eqs.~(\ref{e21}) and (\ref{e21a}), $P^{11}\left( t,t^{\prime }\right)$
and $P^{22}\left( t,t^{\prime }\right)$ are real symmetric functions, while $%
P^{12}\left( t,t^{\prime }\right) =\left( P^{21}\left( t^{\prime },t\right)
\right) ^{\ast }$. 
The contribution of off-diagonal elements, $W^{ab}\left( t,t^{\prime }\right)$, $a
\neq b$,  to Eq.~(\ref{e21}), is subleading at $N\gg1$.
The reason for this is that these elements are Grassmann variables, and upon expanding the exponent in (\ref{e21}), one generates only first-order and mixed second order terms that come with small powers of $N$. Thus, in the main approximation in $N$, contributions coming
from $W^{aa}\left( t,t^{\prime }\right)$ are most important and we
concentrate on them.

To simplify the action and analyze its equations of motion, we have to
decouple the terms $\left( \mathrm{Tr}\left( W\left( t,t^{\prime }\right)
W\left( t^{\prime },t\right) \right) \right) ^{2}$ by one more gaussian
decoupling. To do this we introduce a new diagonal matrix-field $%
Q^{aa}\left( t,t^{\prime}\right)$, $a=1,2$, and use the following identities
\begin{eqnarray}
&&\exp \Big[Ni\int_{-\infty }^{\infty }P^{aa}\left( t,t^{\prime }\right)
W^{aa}\left( t,t^{\prime }\right) W^{aa}\left( t^{\prime },t\right)
dtdt^{\prime }\Big]  \notag \\
&=&\int DQ \exp \Big[-Ni\sum_{ij}^{N}\int_{-\infty }^{\infty }P^{aa}\left(
t,t^{\prime }\right) \Big[\mathrm{Tr}\Big[Q^{aa}\left( t,t^{\prime }\right)
Q^{aa}\left( t^{\prime },t\right)  \notag \\
&&+2Q^{aa}\left( t,t^{\prime }\right) \Psi ^{a}\left( t^{\prime }\right)
\bar{\Psi}^{a}\left( t\right) \Big]\Big]dtdt^{\prime }\Big] Z_{a}\left[ P%
\right]  \notag \\
&=&Z_{a}\left[ P\right] \int DQ\exp \Big[-Ni\sum_{ij}\int_{-\infty }^{\infty
}  \notag \\
&&\times \Big[\mathrm{Tr}\Big[P^{aa}\left( t,t^{\prime }\right) Q^{aa}\left(
t,t^{\prime }\right) Q^{aa}\left( t^{\prime },t\right) \Big]  \notag \\
&&-2\left( -1\right) ^{a-1}\bar{\Psi}^{a}\left( t\right) P^{aa}\left(
t,t^{\prime }\right) Q^{aa}\left( t,t^{\prime }\right) \Psi ^{a}\left(
t^{\prime }\right) \Big] \notag \\
&&dtdt^{\prime }\Big],  \notag \\
&&  \label{e22}
\end{eqnarray}
where $a=1,2$, and we introduced the following notation:
\begin{eqnarray}
&&Z_{a}\left[ P\right]=  \label{e24a} \\
&&\int DQ\exp \Big[iN\sum_{ij}\int_{-\infty }^{\infty }P^{aa}\left(
t,t^{\prime }\right) Q^{aa}\left( t,t^{\prime }\right) Q^{aa}\left(
t^{\prime },t\right) \Big].  \notag
\end{eqnarray}
In Eq.~(\ref{e22}), the new matrices
\begin{equation}
Q\left( t,t^{\prime }\right) =\left(
\begin{array}{cc}
Q^{11}\left( t,t^{\prime }\right) & 0 \\
0 & Q^{22}\left( t,t^{\prime }\right)%
\end{array}%
\right)  \label{e23b}
\end{equation}%
have the following symmetry
\begin{equation}
\bar{Q}\left( t,t^{\prime }\right) =CQ^{T}\left( t,t^{\prime }\right)
C^{T}=Q^{\dagger}\left( t,t^{\prime }\right) .  \label{e23c}
\end{equation}

All these decouplings and notations allow us to write the full partition
function, $Z$, of the model in the form
\begin{equation}
Z=\int \exp \left[ iS\left[ \Psi ,\Psi ^{\dagger},P,Q\right] \right] Z\left[ P%
\right] D\Psi DPDQ,  \label{e25a}
\end{equation}%
where the integrant contains a factor $Z\left[ P\right]$ given by
\begin{equation}
Z\left[ P\right] =Z_{a}\left[ P\right] Z_{b}\left[ P\right] Z_{0}.
\label{e25b}
\end{equation}
In Eq.~(\ref{e25a}), the functional $S\left[ \Psi ,\Psi ^{\dagger},P,Q\right] $ is
given by
\begin{eqnarray}  \label{S-2}
&&S\left[ \Psi ,\Psi ^{\dagger},P,Q\right] =\int_{-\infty }^{\infty }dtdt^{\prime }
\notag \\
&&\times \Big[\sum_{i=1}^{N}\bar{\Psi}_{i}\left( t\right) \big[\delta
_{t,t^{\prime }}\left( i\partial _{t^{\prime }}+\tau _{3}\mu \right)
+2P\left( t,t^{\prime }\right) Q\left( t,t^{\prime }\right) \big]\Psi
_{i}\left( t^{\prime }\right)  \notag \\
&&-N\sum_{i,j=1}^{N}\mathrm{Tr}\left( P\left( t,t^{\prime }\right) Q\left(
t,t^{\prime }\right) Q\left( t^{\prime }t\right) \right)  \label{e25c} \\
&&+\frac{iN}{2J^{2}}\mathrm{Tr}\left( P^{2}\left( t,t^{\prime }\right)
\right) \Big],  \notag
\end{eqnarray}%
where
\begin{equation}
P\left( t,t^{\prime }\right) =\left(
\begin{array}{cc}
P^{11}\left( t,t^{\prime }\right) & 0 \\
0 & P^{22}\left( t,t^{\prime }\right)%
\end{array}
\right) .  \label{e25d}
\end{equation}%
Integrating out the supervectors $\Psi ,\Psi ^{\dagger}$, one obtains, using Eq.~(%
\ref{e25a}), the following formula for the partition function $Z$:
\begin{equation}
Z=\int Z\left[ P,Q\right] DPDQ,  \label{e26}
\end{equation}%
with the integrant $Z\left[ P,Q\right] $ being equal to
\begin{eqnarray}
&&Z\left[ P,Q\right] =\exp \Big[N\int_{-\infty }^{\infty }\int_{-\infty
}^{\infty }dtdt^{\prime }\Big[-\frac{\mathrm{Tr}P^{2}\left( t,t^{\prime
}\right) }{2J^{2}}  \notag \\
&&+\mathrm{Tr}\Big[k\mathrm{\ln }\left[ \delta \left( t-t^{\prime }\right)
\left( i\partial _{t^{\prime }}+\tau _{3}\mu \right) +2P\left( t,t^{\prime
}\right) Q\left( t,t^{\prime }\right) \right]  \notag \\
&&-i\mathrm{Tr}\big[P\left( t,t^{\prime }\right) Q\left( t,t^{\prime
}\right) Q\left( t^{\prime },t\right) \big]\Big]\Big].  \label{e28}
\end{eqnarray}%
Here we introduced a $2\times 2$ matrix, $k$, that differentiates between
bosonic and fermionic superpartners,
\begin{equation*}
k=\left(
\begin{array}{cc}
1 & 0 \\
0 & -1%
\end{array}%
\right) .
\end{equation*}

Presence of the large $N$ in the exponential in Eq.~(\ref{e28}\textbf{) }%
allows one to calculate the integral over $P\left( t,t^{\prime }\right)$ and
$Q\left( t,t^{\prime }\right)$ using the saddle-point method. Minimizing
action $-\ln Z\left[ P,Q\right] $ with respect to the matrices $Q\left(
t,t^{\prime }\right)$ and $P\left( t,t^{\prime }\right)$, we obtain the
following saddle-point equations
\begin{equation}
Q\left( t,t^{\prime }\right) =-ik\left[ \delta \left( t-t^{\prime }\right)
\left( i\partial _{t^{\prime }}+\tau _{3}\mu \right) +2P\left( t,t^{\prime
}\right) Q\left( t,t^{\prime }\right) \right] ^{-1},  \label{e29}
\end{equation}%
\begin{eqnarray}
&&P\left( t,t^{\prime }\right) =-iJ^{2}Q\left( t,t^{\prime }\right) Q\left(
t^{\prime },t\right)  \notag \\
&&+2J^{2}Q\left( t,t^{\prime }\right) k\left[ \left( i\partial _{t}+\tau
_{3}\mu \right) \delta \left( t-t^{\prime }\right)\right.  \notag \\
&&\left.+2P\left( t,t^{\prime }\right) Q\left( t,t^{\prime }\right) \right]
^{-1}.  \notag \\
&&  \label{e30}
\end{eqnarray}%
Using Eq.~(\ref{e29}) we rewrite Eq.~(\ref{e30}) in a simpler form
\begin{equation}
P\left( t,t^{\prime }\right) =iJ^{2}Q\left( t,t^{\prime }\right) Q\left(
t^{\prime },t\right) .  \label{e31}
\end{equation}%
As the next step, substituting Eq.~(\ref{e31}) into Eq.~(\ref{e29}), one
will obtain a closed equation for $Q\left( t,t^{\prime }\right) $
\begin{eqnarray}
Q\left( t,t^{\prime }\right) &=&-ik\Big[\left( i\partial _{t}+\tau _{3}\mu
\right) \delta \left( t-t^{\prime }\right)  \notag \\
&&+2iJ^{2}Q\left( t,t^{\prime }\right) Q\left( t^{\prime },t\right) Q\left(
t,t^{\prime }\right) \Big]^{-1}.  \label{e33a}
\end{eqnarray}
Note that, Eq.~(\ref{e33a}) can also be written in a form of a differential
equation,
\begin{eqnarray}
&&\Big[\left( i\partial _{t}+\tau _{3}\mu \right) Q\left( t,t^{\prime
}\right)  \notag \\
&&+2iJ^{2}\int Q\left( t,t^{\prime \prime }\right) Q\left( t^{\prime \prime
},t\right) Q\left( t,t^{\prime \prime }\right) \Big]  \notag \\
&&\times Q\left( t^{\prime \prime },t^{\prime }\right) dt^{\prime \prime }
=-ik\delta \left( t-t^{\prime }\right).  \label{e34}
\end{eqnarray}%
As the function $Q\left( t,t^{\prime }\right) $ is diagonal, one can solve
Eq.~(\ref{e34}) separately for the fermion and boson parts. At small
energies (large-time limit) and $\mu=0$, one can neglect the first line in Eq.~(\ref{e34}). Assuming that the solutions depend on the time difference, one comes to
the following set of equations 
\begin{eqnarray}
&&2J^{2}\int \left( Q^{F}\left( t-t^{\prime \prime }\right) \right)
^{2}Q^{F}\left( t^{\prime \prime }-t\right) Q^{F}\left( t^{\prime \prime
}-t^{\prime }\right) dt^{\prime \prime }  \notag \\
&=&-\delta \left( t-t^{\prime }\right) ,  \notag \\
&&2J^{2}\int \left( Q^{B}\left( t-t^{\prime \prime }\right) \right)
^{2}Q^{B}\left( t^{\prime \prime }-t\right) Q^{B}\left( t^{\prime \prime
}-t^{\prime }\right) dt^{\prime \prime }  \notag \\
&=&\delta \left( t-t^{\prime }\right) ,  \label{e35a}
\end{eqnarray}%
where $Q^{F}\left( t,t^{\prime }\right) $ and $Q^{B}\left( t,t^{\prime
}\right) $ are fermion and boson parts the matrix $Q\left( t,t^{\prime
}\right)$. The structure of Eqs.~(\ref{e35a}) is similar that of equations
obtained in Ref.~\onlinecite{Sachdev-2015}, although there are small
differences due a fact that we considered here a slightly different model,
Eq.~(\ref{e4}), written in real time.

From Eqs.~(\ref{e31})-(\ref{e34}), we can find the Green's function of
fermions and bosons in energy space 
\begin{equation*}
G(\omega )=\left[ {\omega \mathbb{1}_4-\Sigma (\omega )}\right] ^{-1}
\end{equation*}%
where $ \mathbb{1}_4-$ is a $4$ dimensional identity matrix and $\Sigma (\omega )$ is the Fourier image of the electron/boson
self-energy,
\begin{equation}
\Sigma \left( t-t^{\prime }\right) =-2 J^2 k G^{2}(t-t^{\prime })G(t^{\prime
}-t).  \label{selfenergy}
\end{equation}%
In Eq.~(\ref{selfenergy}) $\Sigma$ and $G$ are $4\times 4$ diagonal
matrices. For fermionic part this relations fully coincide with ones
obtained in Ref. \onlinecite{Sachdev-2015}, while bosonic self energy has
the opposite sign, as it should be. One can see easily that this sign
difference gives the unity partition function $Z\left[ P,Q\right]$ given by
Eqs.~(\ref{e26}), (\ref{e28}). Indeed, writing the derivative of the
logarithm of the partition function $Z$ and using the saddle point equations
(\ref{e29}, \ref{e31}), we obtain
\begin{equation}
\mathbf{-}\frac{\partial }{\partial J}\ln Z\left[ P_{J},Q_{J}\right] =-\frac{%
N}{J^{3}}\mathrm{Tr}P_{J}^{2}\left( t,t^{\prime }\right),  \label{e36}
\end{equation}
where $Q_{J}$ and $P_{J}$ are solutions of the saddle point equations (\ref%
{e29}, \ref{e31}). Using Eqs.~(\ref{e31}), (\ref{e35a}), (\ref{e36}) and
reconstructing the partition function $Z\left[ P_{J},Q_{J}\right]$ from its
derivative we conclude that it is equals one. This confirms that the saddle
point solution does not contradict the supersymmetry. Although study of the
solution of Eqs.~(\ref{e35a}) at arbitrary time is also interesting, we do
not perform it here.


In the scaling low energy limit $\omega <<J$, and zero chemical potential,
the expression for the Green's function $G_{a}(\omega )$ (where $a=1,2$
corresponds to fermions and $a=3,4$ to bosons) has one dimensional time
reparametrization, $t =f(\sigma )$, and emergent $U(1)$ gauge invariance,
defined by Sachdev in Ref.~\onlinecite{Sachdev-2015} for imaginary time:
\begin{eqnarray}
G_{a}(t,t^{\prime }) &=&[f^{\prime }(\sigma )f^{\prime }(\sigma
^{\prime})]^{-1/4}\frac{g(\sigma )}{g(\sigma ^{\prime })}G_{a}(\sigma
,\sigma ^{\prime })  \notag  \label{reparam} \\
\Sigma _{a}(t ,t ^{\prime }) &=&[f^{\prime }(\sigma )f^{\prime }(\sigma
^{\prime})]^{ -3/4}\frac{g(\sigma )}{g(\sigma ^{\prime })}\Sigma _{a}(\sigma
,\sigma ^{\prime }).
\end{eqnarray}%
Here $f(\sigma )$ and $g(\sigma )$ are arbitrary functions. These symmetries
impose strong restrictions on $G$ and $\Sigma $ and lead to the following
asymptotic expression for the Green's function at zero temperature:
\begin{equation*}
G_{1}(t )=\Bigg\{%
\begin{array}{c}
\frac{C e^{3i\pi/4}\sin [\pi /4+\theta ]}{\sqrt{\pi t }},\;\;\;t >>1/J \\
\frac{C e^{-3i\pi/4} \sin [\pi /4+\theta ]}{\sqrt{-\pi t }},\;\;\;\;-t >>1/J,%
\end{array}%
\end{equation*}
with constant $C$. These expressions were first obtained in Ref.~%
\onlinecite{Sachdev-2015}. Therefore at least in the asymptotic regime of
large times (low energies), we do not expect a difference between our
supersymmetric formulation of SYK model and the replica approach to it.
However, at intermediate times, when we can not ignore the kinetic term for
supersymmetric (fermionic) fields in action (\ref{S-2}), a difference may be
essential.


\section{Superbosonization of SYK model}


In Section IV, we explicitly developed a supersymmetry method for
interacting SYK model, which produced non-perturbative results. Remarkably,
in the above-developed approach, the supersymmetry is explicit at the level
of the saddle point equations. These equations are very interesting, and may
potentially provide some more new information about the system behavior at
various energy scales. At the same time, as we can see from saddle point
equations (\ref{e35a}), the fermion and boson sectors of the diagonal matrix
field $Q\left(t,t^{\prime }\right) $ are decoupled. Thus bosons and fermions
do not interfere with each other in this formulation.

Interestingly enough, there is an alternative, conceptually similar but
technically different, way of formulating the SYK model as a supersymmetric $%
\sigma$-model. It is the superbosonization procedure, which will be
developed in this section. We will show that at the level of the saddle
point equations in the superbosonized description, bosonic degrees of
freedom interfere with fermions. This interference effect can be accounted
for analytically. It may potentially become crucial for revealing novel
modes in correlation functions - the advantage of the supersymmetric
approaches as compared to replica and imaginary time methods is that they
allow for controlled analysis of the intermediate time regime.

Consider a function $F(\Phi \otimes {{\Phi }^{\dagger}})$ of the tensor product of
a super-vector $\Phi $ and its conjugate ${{\Phi }^{\dagger}}$ given by Eq.~(\ref%
{e10}). Generally, after ensemble averaging of disordered single-particle
systems, one deals with integrals of type $\int D\Phi D{{\Phi }^{\dagger}}\;F(\Phi
\otimes {{\Phi }^{\dagger}})$. The super-bosonization formula essentially allows
evaluating such a supervector integral to an integral over a supermatrix $Q$%
, where Q has no constraints (unlike direct product $\Phi \otimes {{\Phi }%
^{\dagger}}$).

Being formally exact, the superbosonization approach \cite%
{Efetov-2003,Efetov-2004,Sedrakyan-2017} proved to be very efficient in
producing nonperturbative results for example in the theory of almost
diagonal random matrices \cite{Sedrakyan-2017,
Yevtushenko-2003,Yevtushenko-2004}, where the standard supersymmetry method
was also instrumenal\cite{oleg1,oleg2}. To derive the superbosonized
representation of the SYK model, here we will follow a slightly different
path from the one outlined in Section III. In contrast with Eq.~(\ref{e9}),
wherein the joined fermion-boson action contained two different
Hubbard-Stratonovich fields, $M_F(t)$ and $M_B(t)$ defined in (\ref{e11}) for
fermions and bosons respectively, here we introduce a unique field, $M(t)$%
\cite{footnote}. This procedure is allowed because of the property that the
determinant in the denominator Eq.~(\ref{denominator}) is independent of the
fluctuating Hubbard-Stratonovich field. Then this procedure will lead to to
the action
\begin{eqnarray}  \label{A1}
S&=&\int dt \Big[\sum_{i,a}\Phi^+_{i,a}\big[(i \partial_t + \mu)\delta{ij}-
2 M_{ij}\big]\Phi_{i,a}  \notag \\
&+&\sum_{ijkl, a,b}M_{ij}[J^{-1}]M_{kl}\Big].
\end{eqnarray}
Further, we integrate over the Gaussian fluctuating field, $M$. This
procedure gives the following expression for the action
\begin{eqnarray}  \label{A11}
S&=&\int dt \sum_{i,a}\Phi^+_{i,a}\big[(i \partial_t +\mu)\Phi_{i,a}
\notag \\
&+&\sum_{ijkl, a,b} J_{ijkl} \Phi _{i,a}^{\dagger}(t)\Phi _{j,a}(t)\Phi
_{k,b}^{\dagger}(t)\Phi _{l,b}(t) \Big].
\end{eqnarray}
As the next step, we perform disorder averaging. The integration measure of
random couplings, $J_{ijkl}$, is Gaussian: $\sim e^{-N^3
\sum_{ijkl}J^{\dagger}_{ijkl}J_{ijkl}/8J^2}$ with $J^{\dagger}_{ijkl}=J_{jilk}$.
However, since the couplings have a property of $%
J_{ijkl}=-J_{ilkj}=-J_{kjil}=J_{klij}$, only half of them are independent.
We can select the independent part of couplings $J_{ijkl}$ by using the
ordering of the indexes and choosing $i>k, j>l$ term. Other terms with $i>k,
j<l$, $i<k, j>l$, $i<k, j<l$ are equal to selected one with appropriate
sign. The measure over independent couplings thus becomes
\begin{eqnarray}
\mathcal{W}(J)=e^{-\frac{ N^3}{2 J^2}\sum_{i>k,j>l}|J_{ijkl}|^2}.
\end{eqnarray}
According to the ordering of indices decribed above, the interaction term in
the right-hand-side of (\ref{A11}) is a sum of 4 independent terms $J_{ijkl}$%
, $i>k$, $j>l$:
\begin{eqnarray}  \label{A12}
&&\sum_{a,b}\sum_{ijkl} J_{ijkl} \Phi _{i,a}^{\dagger}(t)\Phi _{j,a}(t)\Phi
_{k,b}^{\dagger}(t)\Phi _{l,b}(t) \\
&=&\sum_{a,b}\sum_{i>k,j>l} 2J_{ijkl} \Big[\Phi _{i,a}^{\dagger}(t)\Phi
_{j,a}(t)\Phi _{k,b}^{\dagger}(t)\Phi _{l,b}(t)  \notag \\
&-& \Phi _{i,a}^{\dagger}(t)\Phi _{l,a}(t)\Phi _{k,b}^{\dagger}(t)\Phi _{j,b}(t)\Big].
\end{eqnarray}
The disorder averaging (i.e., the integration over independent $J_{ijkl}$)
thus produces an interacting theory with action that is similar to the one
in Eq.~(\ref{S}):
\begin{eqnarray}  \label{A2}
&&S=\int dt\Big\{\sum_{i,a}\Phi _{i,a}^{\dagger}\left( t\right) (i\partial
_{t}+\mu )\Phi _{i,a}\left( t\right)  \notag  \label{S-4} \\
&+&\frac{2 iJ^{2}}{ N^{3}}\int dt dt^{\prime }\sum_{aba^{\prime }b^{\prime
}}\sum_{i>k,j>l}  \notag \\
&&\Big[\Phi _{i,a}^{\dagger}(t)\Phi _{j,a}(t)\Phi _{k,b}^{\dagger}(t)\Phi _{l,b}(t)
\notag \\
&\times &\Phi _{j,a^{\prime }}^{\dagger}(t^{\prime })\Phi _{i,a^{\prime
}}(t^{\prime })\Phi _{l,b^{\prime }}^{\dagger}(t^{\prime })\Phi _{k,b^{\prime
}}(t^{\prime }) \\
&-&\Phi _{i,a}^{\dagger}(t)\Phi _{j,a}(t)\Phi _{k,b}^{\dagger}(t)\Phi _{l,b}(t)  \notag
\\
&\times &\Phi _{l,a^{\prime }}^{\dagger}(t^{\prime })\Phi _{i,a^{\prime
}}(t^{\prime })\Phi _{j,b^{\prime }}^{\dagger}(t^{\prime })\Phi _{k,b^{\prime
}}(t^{\prime })\Big] \Big\}.  \notag
\end{eqnarray}
Eq.~(\ref{A2}) is invariant under the supersymmetry transformation $\delta
\chi_i = \epsilon b_i,\; \delta b_i =- \epsilon \chi_i, \; i=1 \cdots N$,
where $\epsilon$ is an infinitesimal Grassmann parameter. The reason for
this is that the building block of the action, namely $\Phi_{i,a}^+
\Phi_{i,a}$, is invariant.

There are two distinct approaches for superbosonization of the SYK model. A
general approach is based on the introduction of identity into the partition
function:
\begin{equation}
1=\int_{{\mathbb{H}}_{n}}dQ_{ia,jb}(t,t^{\prime })\delta \Big(%
Q_{ia,jb}\left( t,t^{\prime }\right) -\Phi _{ia}(t)\Phi _{jb}^{\dagger}(t^{\prime
})\Big)  \label{Q1}
\end{equation}%
Here $Q_{ia,jb}(t ,t^{\prime })$, is a non-local supermatrix. The second,
simpler way would be through introducing
\begin{equation}
1=\int_{{\mathbb{H}}_{n}}dQ_{aa^{\prime }}^{i}(t,t^{\prime })\delta \Big(%
Q_{aa^{\prime }}^{i}\left( t,t^\prime\right) -\Phi _{ia}(t)\Phi _{ia^{\prime
}}^{\dagger}(t^{\prime })\Big),  \label{Q2}
\end{equation}%
imposed by the non-local matrix $Q_{aa^{\prime }}^{i}(t,t^{\prime })$. Here $%
{\mathbb{H}}_{n}$ is the linear space of Hermitian $2n\times 2n$
supermatrices. We recall, that formal sums of formal products $\Phi \otimes {%
{\Phi }^{\dagger}}$, where $\Phi \in U(n,1|n,1)$ and ${\ \Phi }^{\dagger}\in U(n,1|n,1)$
are supervectors, constitute a vector space. This vector space is defined,
up to isomorphism, by the condition that every antisymmetric, bilinear map $%
f:U(n,1|n,1)\times {\bar{U}}(n,1|n,1)\rightarrow {\mathbb{G}}$ determines a
unique linear map $g:U(n,1|n,1)\otimes {\bar{U}}(n,1|n,1)\rightarrow {%
\mathbb{G}}$ with $f(\Phi ,{\Phi ^{\dagger}})=g(\Phi \otimes {{\Phi }^{\dagger}})$. This
implies that if we consider a map, ${\mathcal{F}}:{\mathbb{H}}%
_{n}\rightarrow {\mathbb{G}}$, then the integral $\int D\Phi D{{\Phi }^{\dagger}}%
\;F(\Phi \otimes {{\Phi }^{\dagger}})$ is now well defined. From now on we will
restrict ourselves to the case of maps, ${\mathcal{F}}$, such that the above
integral is convergent.

The delta-function in Eqs.~(\ref{Q1}) and (\ref{Q2}) is a functional defined
as in Ref.~\onlinecite{Sedrakyan-2017}. Namely, for all ${\mathcal{A}}\in {\
\mathbb{H}}_{n}$ the convergent integral $\delta ({\mathcal{A}})=\lim_{\eta
\rightarrow 0}\int_{{\mathbb{H}}_{n}}DB\exp \left\{ i{\text{Str}}[{\mathcal{%
A }}B]-\tilde{\eta}{\text{Str}}[B^{2}]\right\}$, 
taken over ${\mathbb{H}}_{n}$ with flat Berezin measure\cite{berezin}, where symbol "$\mathrm{Str}$" stands for supertrace,
satisfies the condition $\label{delta}\int_{{\mathbb{H}}_{n}}D{\mathcal{A}}
^{\prime }\;{\delta }({\mathcal{A}}^{\prime }-{\mathcal{A}})\equiv 1$.
Moreover, for any map, ${\mathcal{F}}:{\mathbb{H}}_{n}\rightarrow {\mathbb{G}%
}$, that converges exponentially (or faster), the identity ${\mathcal{F}}%
(Q)\equiv \int_{{\mathbb{H}}_{n}}D{\mathcal{A}}{\mathcal{F}}({\mathcal{A}}%
)\delta ({\mathcal{A}}-Q)$ 
always holds.

Using the above expression for the delta-functional in Eq.~(\ref{Q2}), and
inserting the identity to the partition function defined by Eq.~(\ref{S-4}), we obtain an effective action
\begin{eqnarray}
&&S =\int dtdt^{\prime }\Bigg\{\sum_{i}\Big[(i\partial _{t}+\mu )\delta
_{tt^{\prime }}\mathrm{Str}\big[Q^{i}(t,t^{\prime })\big]  \notag
\label{S-5} \\
&&+\sum_{a}\Phi _{i,a}^{\dagger}\left( t\right) B_{aa^{\prime }}^{i}(t,t^{\prime })\Phi
_{i,a^{\prime }}\left( t^{\prime }\right) +\mathrm{Str}\big[B^{i}\left(
t,t^{\prime }\right) Q^{i}(t^{\prime },t)\big]  \notag \\
&&-\eta \mathrm{Str}\big[B^{i}\left( t,t^{\prime }\right)
B^{i}\left( t,t^{\prime }\right) \big]\Big]+\frac{2 iJ^{2}}{ N^{3}}\times  \notag
\\
&&\times\sum_{i>k,j>l}\Big[\mathrm{Str}\big[Q^{i}\left( t,t^{\prime }\right)
Q^{j}(t^{\prime },t)\big]\mathrm{Str}\big[Q^{k}\left( t,t^{\prime }\right)
Q^{l}(t^{\prime },t)\big]  \notag \\
&&-\mathrm{Str}\big[Q^{i}\left( t,t^{\prime }\right) Q^{j}(t^{\prime
},t)Q^{k}\left( t,t^{\prime }\right) Q^{l}(t^{\prime },t)\big] \Big]\Bigg\}.
\end{eqnarray}
We see that the
superfield $\Phi _{i,a}\left( t\right) $ enters into this action only as a
quadratic form with the matrix $B^{i}(t,t^{\prime })$. Therefore, the
integral over superfields $\Phi _{i,a}$ in the partition function can be
exactly evaluated, producing the superdeterminant of $B^{i}(t,t^{\prime })$
in the denominator of the integrand. It is worth to mention that the
supermatrix $B$ should be considered as a matrix by its arguments $%
B^{i}\left( t,t^{\prime }\right) =B_{t,t^{\prime }}^{i}$. The partition
function of the model thus becomes
\begin{equation}
Z=\int \prod_{i}\mathcal{D}B^{i}(t,t^{\prime })\mathcal{D}Q^{i}(t,t^{\prime
})\mathrm{Sdet}[B]\exp \{iS\},  \label{PF1}
\end{equation}
where "$\mathrm{S\det }$" is superdeterminant.

Now, omitting for a while the first two terms in Eq. (\ref{S-5}), we
introduce a notation $\tilde{S}$ for the remaining terms in the expression,
and write is in the form
\begin{eqnarray}
&&\tilde{S} =\lim_{\eta \rightarrow 0}\int_{-\infty }^{\infty }\int_{-\infty
}^{\infty }dtdt^{\prime }\Bigg\{\mathrm{Str}\big[B^{i}\left( t,t^{\prime
}\right) Q^{i}(t^{\prime },t)\big]  \notag  \label{S-6} \\
&&+i\eta \sum_{i}\mathrm{Str}\big[\big(B^{i}(t,t^{\prime })\big)^{2}\big] +%
\frac{2 i J^{2}}{ N^{3}}\times \\
&&\times\sum_{i>k,j>l}\Big[\mathrm{Str}\big[Q^{i}\left( t,t^{\prime }\right)
Q^{j}(t^{\prime },t)\big]\mathrm{Str}\big[Q^{k}\left( t,t^{\prime }\right)
Q^{l}(t^{\prime },t)\big]  \notag \\
&&-\mathrm{Str}\big[Q^{i}\left( t,t^{\prime }\right) Q^{j}(t^{\prime
},t)Q^{k}\left( t,t^{\prime }\right) Q^{l}(t^{\prime },t)\big] \Big]\Bigg\},
\notag
\end{eqnarray}
Then following the method introduced in Ref.~\onlinecite{Sedrakyan-2017}, we
join $Q$ with $B$ by introducing a new supermatrix $\bar{B}=BQ$. Here we
note that the formal sums of Hermitian super-bivectors (product of two
supermatrices, each of them being from the linear space of complex Hermitian
supermatrices ${\mathbb{H}}_{n}$), constitute a vector space $\Lambda ^{2}({%
\mathbb{H}}_{n})$ called the second exterior power of ${\mathbb{H}}_{n}$.
Then, integration over $\bar{B}\in \Lambda ^{2}({\mathbb{H}}_{n})$ decouples
from the partition function and produces a constant
\begin{eqnarray}
C_{n} &=&\int_{\Lambda ^{2}({\mathbb{H}}_{n})}\mathcal{D}\bar{B}\mathrm{Sdet}%
[\bar{B}]\exp \Big\{\int dtdt^{\prime }\mathrm{Str}[\bar{B}(t,t^{\prime })]%
\Big\}.  \notag \\
&&  \label{B2}
\end{eqnarray}
The Berezinian of the transformation $\bar{B}=BQ$ is one. One can see this
by explicitly writing the transformations using the matrix form of $\bar{B}$
(with first indices corresponding to fermion or boson fields). Namely, the
Jacobian of the transformation $\bar{B}_{Bc}=B_{Ba}Q_{ac}$ is $\mathrm{Sdet}%
[Q]$, while for $\bar{B}_{Fc}=B_{Fa}Q_{ac}$ the Jacobian is $1/\mathrm{Sdet}%
[Q]$. As a result of the transformation, these two terms cancel each other
in the product. This happens because the fields $\bar{B}_{Fc}$ and $\bar{B}%
_{Bc}$ always have opposite fermionic parity.

Finally, the partition function $Z$ acquires the form
\begin{equation}
{Z}=\int_{{\mathbb{H}}_{n}}\prod_{i}\mathcal{D}Q^{i}(t,t^{\prime })\frac{1}{%
\mathrm{SDet}[Q^{i}(t,t^{\prime })]}\exp \{i\tilde{S}(Q)\},  \label{PF2}
\end{equation}%
with the action
\begin{widetext}
\begin{eqnarray}
\tilde{S}(Q) =\int dtdt^{\prime }\Bigg\{\sum_{i}\mathrm{Str}\big[(i\partial
_{t}+\mu )\delta _{tt^{\prime }}Q^{i}(t,t^{\prime })\big]  \label{S-7}
&+&\frac{2 i J^{2}}{ N^{3}}\sum_{i>k,j>l}\Big[\mathrm{Str}\big[Q^{i}\left(
t,t^{\prime }\right) Q^{j}(t^{\prime },t)\big]\mathrm{Str}\big[Q^{k}\left(
t,t^{\prime }\right) Q^{l}(t^{\prime },t)\big]
\notag \\
&-&\mathrm{Str}\big[Q^{i}\left( t,t^{\prime }\right) Q^{j}(t^{\prime
},t)Q^{k}\left( t,t^{\prime }\right) Q^{l}(t^{\prime },t)\big] \Big]\Bigg\},
\notag \\
\end{eqnarray}
It is important now to observe that the $2\times 2$ matrix Green's function ${\mathcal G}_i^{ab}(t,t')$ of two-component superfields $\Phi_{ia}(t)$ and $\Phi^{\dagger}_{ib}(t')$ (that contains the fermion propagator in its fermion-fermion block) is
	equal to the vacuum average of the superbosonization matrix field, $\langle Q^i_{ab}\rangle$. Indeed, using the identity Eq.~(\ref{Q2}), one can introduce $Q_i^{ab}(t,t')$ under the integral and obtain
\bea
\label{PhiQ}
{\mathcal G}_i^{ab}(t,t')=-i\langle\Phi_{ia}(t)\Phi^{\dagger}_{ib}(t')\rangle
&=&-i\int {\cal D}\Phi \; \Phi_{ia}(t)\Phi^+_{ib}(t') \exp\{i S\}
\nn\\
 &=& -i\int {\cal D}Q  Q_{{a b}}^i(t,t') exp\{i\tilde{S}(Q)\}
 =-i\langle Q_{ab}^i(t,t')\rangle. \nn
\ena

The functional $\mathrm{SDet}[Q_{i}(t ,t ^{\prime })]$ in the
denominator of the expression (\ref{PF2}) for ${Z}$ should be understood as the
superdeterminant of the supermatrix $Q_{aa^{\prime }}^{i}(t,t^{\prime })$,
which acts linearly in the continuous space of time $t$. Namely, arguments $
t,t^{\prime }$ should be considered as matrix indexes. One can incorporate
the pre-exponent $1/\mathrm{SDet}[Q^{i}(t ,t^{\prime })]$ into the
effective action, $S_{\text{eff}}$, that can be written as
\begin{eqnarray}
&&S_{\text{eff}} =\int dt dt^{\prime }\Bigg\{\sum_{i}\mathrm{Str}\big[%
(i\partial _{t}+\mu )\delta _{tt^{\prime }}Q^{i}(t,t^{\prime })\big]
+\frac{2 iJ^{2}}{ N^{3}}\sum_{i>k,j>l}\Big[\mathrm{Str}\big[Q^{i}\left(
t,t^{\prime }\right) Q^{j}(t^{\prime },t)\big]\mathrm{Str}\big[Q^{k}\left(
t,t^{\prime }\right) Q^{l}(t^{\prime },t)\big]  \notag \qquad \\
&-&\mathrm{Str}\big[Q^{i}\left( t,t^{\prime }\right) Q^{j}(t^{\prime
},t)Q^{k}\left( t,t^{\prime }\right) Q^{l}(t^{\prime },t)\big] \Big]
+i\sum_{i}\mathrm{Str}[\log Q^{i}](t,t^{\prime })\delta (t-t^{\prime })
\Bigg\}.  \label{S-8}
\end{eqnarray}
Here $\log Q^i$ should be understood as the formal series
$\log Q^i=(Q^i-1)+ \frac{1}{2}(Q^i-1)*(Q^i-1)+\frac{1}{3}(Q^i-1)*(Q^i-1)*(Q^i-1)+\cdots$, where the symbol $*$
stands for the convolution product  $[A*B](t,t')=\int dt'' A(t,t'') B(t'',t')$.

\end{widetext}

Now let us analyze the equation of motion of the field $Q^{i}(\tau ,\tau
^{\prime })$ and compare it with the analysis performed in Section III. The
crucial point is that we have additional $\log Q^{i}(t ,t^{\prime})$ term,
which can contribute in the asymptotic analysis. From Eq.~( \ref{Q2}) we see
that $\left\langle Q^{i}\left( t,t^{\prime }\right) \right\rangle
=\left\langle \Phi _{i}(t)\Phi _{i}^{\dagger}\left( t^{\prime }\right)
\right\rangle $ gives the Green function ${\mathcal{G}}^{i}(t,t^{\prime })$
and its asymptotic behavior at large time scale $t\rightarrow \infty $ is
defined by the equation of motion for matrix field $Q^{i}(t,t^{\prime})$:
\begin{eqnarray}
&&\frac{\delta S_{\text{eff}}}{\delta Q^{l}(t,t^{\prime })} =0=(i\partial
_{t}+\mu )\delta _{tt^{\prime }} +i\left[ Q^{l}\left( t.t^{\prime }\right) %
\right] ^{-1}  \notag  \label{SP} \\
&+&\frac{2iJ^{2}}{ N^{3}}\sum_{i>k,j>l}\mathrm{Str}\left[ Q^{i}\left(
t,t^{\prime }\right) Q^{j}\left( t^{\prime },t\right) \right]
Q^{k}(t,t^{\prime }) \\
&-&\frac{2iJ^{2}}{ N^{3}}\sum_{i>k,j>l} Q^{i}\left( t,t^{\prime }\right)
Q^{j}\left( t^{\prime },t\right) Q^{k}(t,t^{\prime }).  \notag
\end{eqnarray}%
This equation shows, that the solutions can be independent of index, $i$, and
therefore we drop it. Putting now $-i\langle Q(t,t^{\prime
})\rangle\equiv{\mathcal{G}}(t,t^{\prime })$ into the Eq.(\ref{SP}), setting $\mu=0$, and
using ${\mathcal{G}}\left( t,t^{\prime }\right)=[ (i\partial _{t})
\delta _{tt^{\prime }}-iK \left( t,t^{\prime }\right)]^{-1}$ with $K\left(
t,t^{\prime}\right)$ being the self-energy, at the large time scale we
obtain
\begin{eqnarray}
K \left( t,t^{\prime }\right) &=&2 J^{2}\mathrm{Str}[\mathcal{G}(t,t^{\prime
})\mathcal{G}(t^{\prime },t)]\mathcal{G}(t,t^{\prime }) \\
&-&2 J^{2} \mathcal{G}(t,t^{\prime })\mathcal{G}(t^{\prime },t)\mathcal{G}%
(t,t^{\prime }).  \notag  \label{SP-2}
\end{eqnarray}
In saddle point approximation, and due to supersymmetry, we expect that the
fermion-fermion (F) and boson-boson (B) entries of the Green's function are
equal: ${\mathcal{G}}_F(t,t^{\prime })=\mathcal{G}_B(t,t^{\prime })$. The
implication of this fact is that that $\mathrm{Str}[\mathcal{G}(t,t^{\prime})\mathcal{G}(t^{%
\prime },t)]=0$, which leads to the relation between the self energy and
Green's functions for fermions and bosons
\begin{eqnarray}
K\left( t,t^{\prime }\right)=- 2 J^{2} {\mathcal{G}}(t,t^{\prime })\mathcal{G%
}(t^{\prime },t)\mathcal{G}(t,t^{\prime }).  \notag  \label{SP-3}
\end{eqnarray}
We note similarity with Eq.~(\ref{selfenergy}) and the similar relation for
fermions obtained using the replica approach. Hence, \emph{at large time
scales} our supersymmetric model reproduces the same asymptotics for the
Green's function as the replica method provides. However, at intermediate
times suppersymmetric action is essentially different from the replica field
theory and we expect that this method will provide new results at
intermediate time scales. In order to see this we rewrite the supertrace
over the supermatrices $Q_i$ in the interaction terms of the action $S_{eff}$%
, defined by (\ref{S-8}), using fermion-boson (FB) and boson-fermion (BF)
components of the supermatrices:
\begin{eqnarray}  \label{boson correction}
&&\mathrm{Str}\big[Q^{i}\left( t,t^{\prime }\right) Q^{j}(t^{\prime },t)\big]%
=Q^{i}_{BF}\left( t,t^{\prime }\right) Q^{j}_{FB}(t^{\prime },t)\qquad \qquad
\notag \\
&-&Q^{i}_{FB}\left( t,t^{\prime }\right) Q^{j}_{BF}(t^{\prime
},t)-Q^{i}_{FF}\left( t,t^{\prime }\right) Q^{j}_{FF}(t^{\prime },t)  \notag
\\
&+&Q^{i}_{BB}\left( t,t^{\prime }\right) Q^{j}_{BB}(t^{\prime },t).
\end{eqnarray}
Similarly for $\mathrm{Str}\big[Q^{k}\left( t,t^{\prime }\right)
Q^{l}(t^{\prime },t)\big]$ part and $\mathrm{Str}\big[Q^{i}\left(
t,t^{\prime }\right) Q^{j}(t^{\prime },t)Q^{k}\left( t,t^{\prime }\right)
Q^{l}(t^{\prime },t)\big] $. BF and FB components of supermatrices are
Grassmann variables and the integration over them is easily performed. It
will produce separate actions for fermions and bosons of the form of Eq.~(%
\ref{S-8}) and additional pre-exponential mixed polynomials from the BB and
FF components. Appearance of these mixed polynomials is a result of the
formally exact supersymmetric approach, and these terms are not captured
within the replica approach. At large time scales, they have subleading
contribution to the correlation function but will have essential
contribution in the intermediate, finite time region. This fact is a major
advantage of the supersymmetric method. More detailed and complete analysis
of this effects is a subject of future investigations.

The supersymmetry method can also be used to study nonperturbative effects in general SYK$_q$ models. 
The spectral correlators in SYK$_q$ models and their deviation from RMT are studied in  Refs.~\onlinecite{Garsia-Garsia-2016, Jia-2019}. It appears that a small number of long-wavelength modes, which can be parameterized via Q-Hermit orthogonal polynomials, describe the deviation.  Moreover, the SYK model with Majorana fermions is more straightforward and can be formulated as a sigma-model [\onlinecite{Altland-2017,Jia-2019}]. The analysis of two-point spectral correlators in two-loop order and shows corrections
to RMT, whose lowest order term corresponds to scale fluctuations in
good agreement with numerical results \cite{Jia-2019} . However, the question remains about other loop terms in the loop expansion and how they should (not) contribute. In general, the range of validity of the loop expansion remains open and is expected to be detectable from the supersymmetry method.  Another obvious open problem is the model at $q=2$. In this case, one should expect Poisson statistics for the spectral correlation. This is in contrast to the replica field theory which suggests an RMT behavior with a significant Thouless scale as shown in Ref.~ \onlinecite{Jia-2019}.

We expect a more straightforward understanding of the RMT structure of the SYK$_q$ model. The superbosonization technique is well developed \cite{Sedrakyan-2017} and can provide exact and nonperturbative results.
In our supersymmetric formulation, there are additional bosonic modes that interact with the original fermions. In the expression (\ref{boson correction}), we present an example of such terms. Just long-wavelength modes of these bosons have the potential of solving the problem of the scale of Thouless transition universally for $q \leq 4$.

Another advantage of the superbosonized $\sigma$-model representation
described above is that it is efficient for computation of correlation
functions. The procedure, described in Ref.~\onlinecite{Sedrakyan-2017}
consists of

1) Diagonalization of $m\times m$ supermatrix field, $Q$, as $Q=UQ_{diag}V$
with diagonalization matrices $U\in U(m\vert m)$ and $V \in U(m\vert
m)/U^{2m} (1)$ restricted to the unitary supergroup and its subspace with
removed phases.

2) After the diagonalization of the supermatrix $Q$, one can integrate over $%
Q$ by integrating over its boson-boson eigenvalues in the interval $\mathbb{R%
}\equiv \{-\infty,\infty\}$, while the integration over the fermion-fermion
eigenvalues should be performed in the interval $i\mathbb{R}%
\equiv\{-i\infty,i\infty\}$.

We see that this procedure significantly reduces the number of integrations
one has to perform to calculate correlation functions within superbosonized
representation.


\section{Conclusions and outlook}


Despite being a standard tool for nonperturbative calculations in disordered
and chaotic systems, the supersymmetric sigma model has rather poorly been
understood for interacting systems. Historically, it was believed that the
Hubbard-Stratonovich decoupling of the interaction Hamiltonian would not
help to develop a supersymmetric description of the partition function of
the model. The reason is that one has to introduce two \emph{different }
Hubbard-Stratonovich bosonic fields, $M_{1}$ and $M_{2}$, to decouple
interaction terms both in the numerator and the denominator of the
expression for any correlation function. It was believed, for about $40$
years, that supersymmetric $\sigma $ model representation of interacting
systems is impossible because fluctuating $M_{1}$ and $M_{2}$ fields are
independent. And therefore, supersymmetry cannot become manifest in a theory
that is disordered, interacting, and dynamical.

In this work, we have challenged this belief and have developed a rigorous supersymmetric $\sigma $-model framework for
interacting disordered systems. The idea that helps to overcome the
abovementioned problem of independence of fluctuating fields $M_{1}$ and $%
M_{2}$ is the following. The partition function of the system is calculated
by the functional integration of an exponentiated action functional over the
space of dynamical field configurations. We showed that for $(0+1)$
dimensional systems, such as quantum dots, the Hubbard-Stratonovich field in
the denominator could be gauged out. It can also be reintroduced back to
guarantee supersymmetry. In order to derive basic formulas of the
supersymmetry method, we have introduced a new version of the SYK model. In
contrast to the previous versions, the model is time-reversal invariant. One
of the main achievements of this paper is that we have given a
supersymmetric $sigma$-model description of the SYK model. We have also
developed its superbosonized description, where the functional integral is
taken over unconstrained dynamical supermatrix fields representing
collective many-body excitations.



It is now a conventional wisdom\cite{Kitaev-2015} that the SYK model
exhibits many-body chaotic properties at all time scales. At short times,
chaos shows up in exponentially decaying correlations as manifested in
out-of-time correlation functions \cite%
{Larkin-1969,Aleiner-2016,Stephen-Maldacena-2016,Maldacena-3-2016,Bagrets-2017,Kitaev-2017}%
. At large time scales, chaos manifests itself in a random matrix ensemble
due to quantum energy level repulsion \cite{Zhuang-2017}. However, the
nature of the transition region from non-ergodic to ergodic regimes remains
unclear. Moreover, the physics of non-ergodic states is not yet fully
understood, and "dirty" metals represent an excellent physically motivated
playground for such studies. Here, an important development was made in Ref.~%
\onlinecite{Miklitz}, where the theoretical description of nonergodic
extended states in a modified SYK model was put forward. The problem of
finding the ergodic (Thouless) time in SYK model was considered in Ref.~%
\onlinecite{Altland-2017}, where the questions regarding the nature of the
relaxation modes, their classification by certain effective quantum numbers,
as well as the density of states, were addressed. An important correlation
function, capable of detecting chaotic properties of the SYK model, is the
spectral number variance $\Sigma _{2}(\epsilon )$. It represents the
statistical variation in the number of many-body levels contained in an
energy window of width $E$. The variance $\Sigma _{2}(\epsilon )$ was
studied in Ref.~\onlinecite{Garsia-Garsia-2016}, where a deviation from the
random matrix ensemble prediction was reported. This deviation demonstrates
the possible breakdown of ergodicity, and this is one of the interesting
points that can be investigated further using superbosonization.

The spectral form factor considered in Ref.~\onlinecite{Cotler-2017},
representing the Fourier transform of the energy-dependent spectral
two-point correlation function, $R_{2}(\epsilon )$, is yet another quantity
of interest. While the longtime profile of it showed a ramp structure
characteristic for random matrix theory ensembles, universal deviations from
random matrix theory were observed for shorter times (see Refs.~\onlinecite
{Li-2017,Hunter-Jones-2017,Cotler-2017-2,delCampo-2017} for related
studies). Density-density correlators were studied in Ref.~%
\onlinecite{Altland-2017} within the replica approach describing the quantum
chaotic dynamics of the SYK model at large times. It was observed that there
are non-ergodic collective modes, which relax in some time interval and
become ergodic states by entering into the longtime regime. The latter modes
can be described using the random matrix theory. These interesting modes
share similar properties with the diffusion modes of dirty metals and have
quantum numbers which have been identified as the generators of the Clifford
algebra \cite{Bagrets-2017}. There, each of the $2N$ different products
formed from $N$ Majorana operators represents a mode.

Here we propose that the superbosonization approach to the SYK model will
open new possibilities to study intermediate time regions and reveal new
aspects of chaotic properties. In particular, it would be fascinating to (i)
calculate one-point correlation function $\langle \rho (E)\rangle $ (the
density of states) in superbosonized representation of SYK model and compare
it with the universal random matrix prediction; (ii) calculate the two-point
correlation function, $\langle \rho (E)\rho (E^{\prime })\rangle $, in SYK
model using its superbosonized representation and compare it with numerical
calculations in Refs.~\onlinecite{Cotler-2017, Altland-2017}; (iii) reveal
the role of bosonic excitations presented in superbosonized representation
and to detect their behavior at short times.

Systematic deviations from the random matrix predictions, for sufficiently
well-separated eigenvalues, imply that the model is not ergodic at short
times. The point of departure from the results of random matrix theory
increases with $N$, which is an indication of having a Thouless energy scale
\cite{Altshuler-1988,Braun-1995,Bertrand-2016, Benet-2001} in the system.
Detection of Thouless time within a superbozonized approach is yet another
exciting project. It would be also interesting to calculate moments of the
spectral density within the supersymmetric sigma model approach.

On another front, it is well-known that Anderson localization can be avoided
under certain conditions for disorder potential supporting long-range\cite%
{moura, Izrail, Garcia} or short-range\cite{SDS,Flores,Dunlap,PP,
TS1,TS2,TS3} correlations in low dimensions. It is thus very interesting to
investigate the effect of introducing correlations to the disordered
interaction constant, $J_{ijkl}$. We expect that such an analysis can also
be performed using the technique outlined in the present work.\newline

\acknowledgments We would like to thank Alexander Altland, Alex Kamenev, Subir Sachdev, Jacobus J. M. Verbaarschot, and Chenan Wei for discussions and valuable comments. 
The research was supported by startup funds from UMass Amherst (T.A.S), by
Deutsche Forschungsgemeinschaft (Project~FE~11/10-1), and by the Ministry of
Science and Higher Education of the Russian Federation in the framework of
Increase Competitiveness Program of NUST \textquotedblleft
MISiS\textquotedblright (Nr.~K2-2017-085") (KBE).


\begin{thebibliography}{999}
\bibitem{Kitaev-2015} A. Kitaev,
Talks at the KITP on April 7th and May 27th (2015).

\bibitem{Sachdev-2015} S. Sachdev, \textit{Bekenstein-Hawking Entropy and
Strange Metals}, Phys. Rev. X \textbf{5}, 041025 (2015).

\bibitem{Maldacena-2016} J. Maldacena and D. Stanford, \textit{Remarks on
the Sachdev-Ye-Kitaev model}, Phys. Rev. D94, 106002 (2016).

\bibitem{agd} A.A. Abrikosov, L.P. Gorkov, and I.E. Dzyaloshinskii, \textit{%
Methods of quantum field theory in statistical physics}, Prentice Hall, New
York (1963).

\bibitem{EA} S.F. Edwards and P.W. Anderson, \textit{Theory of spin glasses}%
, J. Phys. F \textbf{5}, 965 (1975).

\bibitem{LVK} L. V. Keldysh, \textit{Diagram Technique for Nonequilibrium
Processes}, Zh. Eksp. Teor. Fiz. \textbf{47}, 1515 (1964); [Sov. Phys. JETP
\textbf{20}, 1018 (1965)].

\bibitem{Schwinger} J. Schwinger, \textit{Brownian Motion of a Quantum
Oscillator}, J. Math. Phys. \textbf{2}, 407 (1961).

\bibitem{Feynman} R. P. Feynman and F. L Vernon Jr., \textit{The theory of a
general quantum system interacting with a linear dissipative system}, Ann.
Phys. \textbf{24}, 118 (1963).

\bibitem{AK} A. Kamenev, \textit{Field Theory of Non-Equilibrium Systems,}
Cambridge University Press (2012).

\bibitem{Efetov-1983} K.~B.~Efetov, \textit{Supersymmetry and theory of
disordered metals}, Adv. Phys. \textbf{32 }, 53 (1983).

\bibitem{Efetov-1997} K.~B.~Efetov, \textit{Supersymmetry in Disorder and
Chaos}, (Cambridge University Press, Cambridge, 1997).

\bibitem{finkelstein} A.M. Finkelstein, \textit{Influence of Coulomb
interaction on the properties of disordered metals,} Zh. Eksp. Teor. Fiz.
\textbf{84}, 168 (1983) [Sov. Phys. JETP, \textbf{57}, 97 (1983)]

\bibitem{Verbaarschot-1985} J.~J.~M.~Verbaarschot, H.~A.~Weidenmuller, and
M.~R.~Zirnbauer, \textit{Grassmann integration in stochastic quantum
physics: The case of compound-nucleus scattering, }Phys. Rep. \textbf{129},
367 (1985).

\bibitem{Sachdev-1993} S. Sachdev and J. Ye. \textit{Gapless spin-fluid
ground state in a random qantum Heisenberg magnet.} Phys. Rev. Lett.,
\textbf{70}, 3339 (1993).

\bibitem{French} J. B. French, S. S. M. Wong, Phys. Lett. B 33, 449 (1970);
J. B. French, S. S. M. Wong, Phys. Lett. B 35, 5 (1971).
\bibitem{Bohigas} O. Bohigas, J. Flores, Phys. Lett. B 34, 261 (1971); 35, 383 (1971).

\bibitem{Jensen-2016} K. Jensen, \textit{Chaos in AdS2 Holography}, Phys.
Rev. Lett. \textbf{117}, 111601 (2016), arXiv:1605.06098 [hep-th].

\bibitem{Maldacena-2-2016} J. Maldacena, D. Stanford, and Z. Yang,
"Conformal symmetry and its breaking in two dimensional Nearly
Anti-de-Sitter space", PTEP 2016, 12C104 (2016), arXiv:1606.01857 [hep-th].

\bibitem{Engelsoy-2016} J. Engelsoy, T. G. Mertens, and H. Verlinde, "An
investigation of AdS2 backreaction and holography", JHEP 07, 139 (2016),
arXiv:1606.03438 [hep-th].

\bibitem{Larkin-1969} A. I. Larkin and Yu. N. Ovchinnikov, \textit{\
Quasiclassical Method in the Theory of Superconductivity}. Sov. Phys. JETP,
\textbf{28}(6), 1200 (1969).

\bibitem{Aleiner-2016} I. L. Aleiner, L. Faoro, and L. B. Ioffe, \textit{%
Microscopic model of quantum butterfly effect: Out-of-time-order correlators
and traveling combustion waves}, Annals of Physics, \textbf{375}, 378 (2016).

\bibitem{Stephen-Maldacena-2016} J. Maldacena, S.H. Shenker, and D.
Stanford, \textit{A bound on chaos}. Journal of High Energy Physics, \textbf{%
2016}, 106 (2016).

\bibitem{Maldacena-3-2016} J. Maldacena, D. Stanford, and Z. Yang, \textit{\
Conformal symmetry and its breaking in two-dimensional nearly anti-de Sitter
space}, Progress of Theoretical and Experimental Physics, \textbf{2016 }
(12):12C104, (2016).

\bibitem{Jakiw-1985} R. Jackiw, \textit{Lower Dimensional Gravity}, Nucl.
Phys. B252, 343 (1985).

\bibitem{Teitelbom-1983} C. Teitelboim, \textit{Gravitation and Hamiltonian
Structure in Two Space-Time Dimensions}, Phys. Lett. B 126, 41 (1983).

\bibitem{Almheiri-2015} A. Almheiri and J. Polchinski, \textit{Models of
AdS2 backreaction and holography}, JHEP 11, 014 (2015).

\bibitem{Kitaev-2017} A. Kitaev and S. J. Suh, \textit{The soft mode in the
Sachdev-Ye-Kitaev model and its gravity dual.} J. High Energ. Phys. 2018,
183 (2018).

\bibitem{Bagrets-2017} D. Bagrets, A. Altland, and A. Kamenev, \textit{%
Power-law out of time order correlation functions in the SYK model,} Nucl.
Phys. B \textbf{921} , 727 (2017).

\bibitem{altland-2019prl} A. Altland, D. Bagrets, and A. Kamenev, \textit{%
Sachdev-Ye-Kitaev Non-Fermi-Liquid Correlations in Nanoscopic Quantum
Transport}, Phys. Rev. Lett. 123, 226801 (2019).

\bibitem{abk-2019} A. Altland, D. Bagrets, and A. Kamenev,\textit{Quantum
Criticality of Granular Sachdev-Ye-Kitaev Matter}, Phys. Rev. Lett. 123,
106601 (2019).

\bibitem{lunkin} A. V. Lunkin, K. S. Tikhonov, and M. V. Feigel'man \textit{%
Sachdev-Ye-Kitaev Model with Quadratic Perturbations: The Route to a
Non-Fermi Liquid}, Phys. Rev. Lett. 121, 236601 (2018).

\bibitem{bulucheva} Ksenia Bulycheva, \textit{A note on the SYK model with
complex fermions}, JHEP 1712, 069 (2017).

\bibitem{Deutsch-1991} J. Deutsch, \textit{Quantum statistical mechanics in
a closed system,} Physical Review A \textbf{43},2016 (1991).

\bibitem{Srednicki-1994} M. Srednicki, \textit{Chaos and quantum
thermalization,} Phys.l Rev. E \textbf{50}, 888 (1994).

\bibitem{Altland-2017} A. Altland, and D. Bagrets, \textit{Quantum
ergodicity in the SYK model}, Nucl. Phys. B 930, 45-68 (2018).

\bibitem{Haque-1711} M. Haque and P. A. McClarty, \textit{Eigenstate
Thermalization Scaling in Majorana Clusters: from Chaotic to Integrable
Sachdev-Ye-Kitaev Models}, Phys. Rev. B 100, 115122 (2019).

\bibitem{Liu-1709} C. Liu, X. Chen, and L. Balents, \textit{Quantum
entanglement of the Sachdev-Ye-Kitaev models}, Phys. Rev. B 97, 245126
(2018).

\bibitem{yichen} Yichen Huang and Yingfei Gu, \textit{Eigenstate
entanglement in the Sachdev-Ye-Kitaev model}, Phys. Rev. D 100, 041901(R)
(2019).


\bibitem{Garcia-Garcia-2018} A. M. Garcia-Garcia, and M. Tezuka, \textit{%
Many-Body Localization in a finite-range Sachdev-Ye-Kitaev model}, Phys.
Rev. B 99, 054202 (2019).

\bibitem{Danshita-2016} I. Danshita, M. Hanada, and M. Tezuka, \textit{%
Creating and probing the Sachdev-Ye-Kitaev model with ultracold gases:
Towards experimental studies of quantum gravity}, Prog. Theor. Exp. Phys.
083 I 01 (2017).

\bibitem{You-2017} Y.-Z. You, A. W. W. Ludwig, and C. Xu, \textit{%
Sachdev-Ye-Kitaev Model and Thermalization on the Boundary of Many-Body
Localized Fermionic Symmetry Protected Topological States}, Phys. Rev. B 95,
115150 (2017) .

\bibitem{Garsia-Garsia-2016} A. M. Garcia-Garcia and J. J. M. Verbaarschot,
\textit{Spectral and thermodynamic properties of the Sachdev-Ye-Kitaev model}%
, Phys. Rev. D 94, 126010 (2016).


\bibitem{Cotler-2017} J. S. Cotler, G. Gur-Ari, M. Hanada, J. Polchinski, P.
Saad, S. H. Shenker, D. Stanford, A. Streicher, and M. Tezuka, \textit{Black
Holes and Random Matrices}, JHEP 05, 118 (2017).

\bibitem{khveshchenko-1705} D.V. Khveshchenko, \textit{Thickening and
sickening the SYK model}, SciPost Phys. 5, 012 (2018).

\bibitem{khveshchenko-1805} D.V. Khveshchenko,
Condens. Matter 3(4), 40 (2018).

\bibitem{Berkooz-2017} M. Berkooz, P. Narayan, M. Rozali, and J. Simon,
\textit{Higher dimensional generalizations of the SYK model}, JHEP 01, 138
(2017).

\bibitem{Gu-1609} Y.i Gu, X.-L. Qi, D. Stanford, \textit{Local criticality,
diffusion and chaos in generalized Sachdev-Ye-Kitaev models}, J. High Energ.
Phys. 2017, 125 (2017).

\bibitem{Banerjee-2017} S. Banerjee and E. Altman, \textit{Solvable model
for a dynamical quantum phase transition from fast to slow scrambling},
Phys.Rev.B 95, 134302 (2017).

\bibitem{Jian-2017} S.-K. Jian and H. Yao, \textit{Solvable
Sachdev-Ye-Kitaev Models in Higher Dimensions: From Diffusion to Many-Body
Localization}, Phys. Rev. Lett. 119, 206602 (2017).

\bibitem{Haldar-1703} A. Haldar and V. B. Shenoy, \textit{Strange
half-metals and Mott insulators in Sachdev-Ye-Kitaev models}, Phys. Rev. B
98, 165135 (2018).

\bibitem{Banerjee-1710} A. Haldar, S. Banerjee, and V. B. Shenoy, \textit{%
Higher-dimensional Sachdev-Ye-Kitaev non-Fermi liquids at Lifshitz
transitions}, Phys. Rev. B 97, 241106(R) (2018).

\bibitem{Jian-2-2017} C.-M. Jian, Z. Bi, and C. Xu, \textit{Model for
continuous thermal metal to insulator transition}, Phys. Rev. B 96, 115122
(2017).

\bibitem{Gu-2017} Y. Gu, A.Lucas, X.-L. Qi, \textit{Energy diffusion and the
butterfly effect in inhomogeneous Sachdev-Ye-Kitaev chains}, SciPost Phys.
2, 018 (2017).

\bibitem{Song-2017} X.-Y. Song, C.-M. Jian, and L. Balents, \textit{Strongly
Correlated Metal Built from Sachdev-Ye-Kitaev Models}, Phys. Rev. Lett. 119,
216601 (2017).

\bibitem{Chen-2017} X. Chen, R. Fan, Y. Chen, H. Zhai, and P. Zhang, \textit{%
Competition between Chaotic and Nonchaotic Phases in a Quadratically Coupled
Sachdev-Ye-Kitaev Model}, Phys. Rev. Lett. 119, 207603 (2017).

\bibitem{Zhang-2017} P. Zhang, \textit{Dispersive Sachdev-Ye-Kitaev model:
Band structure and quantum chaos}, Phys. Rev.B96, 205138 (2017).

\bibitem{Cai-2018} W. Cai, X.-H. Ge, and G.-H. Yang, \textit{Diffusion in
higher dimensional SYK model with complex fermions}, JHEP 01, 076 (2018).

\bibitem{Zhong-1803} Y.Zhong, \textit{Periodic Anderson model meets
Sachdev-Ye-Kitaev interaction: A solvable playground for heavy fermion
physics}, J. Phys. Commun. 2 095014 (2018).

\bibitem{Mondal-1801} S. Mondal, \textit{Super-maximal chaos and instability}%
, arXiv:1801.09669.

\bibitem{Dai-1802} X. Dai, S.-K. Jian, H. Yao, \textit{Global phase diagram
of the one-dimensional Sachdev-Ye-Kitaev model at finite}, Phys. Rev. B 100,
235144 (2019).


\bibitem{Zirnbauer-1996} M.R. Zirnbauer, \textit{Supersymmetry for systems
with unitary disorder: circular ensembles}, J. Phys. A \textbf{29}, 7113
(1996).

\bibitem{Zirnbauer-1999} M. R. Zirnbauer, \textit{Another critique of the
replica trick}, ArXiv:cond-mat/9903338.

\bibitem{Fu-2017} W. Fu, D. Gaiotto, J. Maldacena, and S. Sachdev, \textit{%
Supersymmetric Sachdev-Ye-Kitaev models}, Phys. Rev. D 95, 026009(2017);
Erratum Phys. Rev. D 95, 069904 (2017).

\bibitem{Fendley-2003} P. Fendley, K. Schoutens, and J. de Boer, \textit{%
Lattice Models with N=2 Supersymmetry}, Phys. Rev. Lett. 90, 120402 (2003).

\bibitem{Fendley-2003-2} P. Fendley, B. Nienhuis, and K. Schoutens, \textit{%
Lattice fermion models with supersymmetry}, J. Phys. A 36, 12399 (2003).

\bibitem{Fendley-2005} P. Fendley and K. Schoutens, \textit{Exact Results
for Strongly Correlated Fermions in 2+1 Dimensions}, Phys. Rev. Lett. 95,
046403 (2005).

\bibitem{Huijse-2008} L. Huijse, J. Halverson, P. Fendley, and K. Schoutens,
\textit{Charge Frustration and Quantum Criticality for Strongly Correlated
Fermions}, Phys. Rev. Lett. 101, 146406 (2008).

\bibitem{Huijse-2010} L. Huijse and K. Schoutens, \textit{Supersymmetry,
lattice fermions, independence complexes and cohomology theory}, Adv. Theor.
Math. Phys. 14, 643 (2010).

\bibitem{Huijse-2011} L. Huijse, N. Moran, J. Vala, and K. Schoutens,
\textit{Exact ground states of a staggered supersymmetric model for lattice
fermions}, Phys. Rev. B 84, 115124 (2011).

\bibitem{Huijse-2012} L. Huijse, D. Mehta, N. Moran, K. Schoutens, and J.
Vala, \textit{Supersymmetric lattice fermions on the triangular lattice:
superfrustration and criticality}, New J. Phys. 14, 073002 (2012).

\bibitem{Anninos-2016} D. Anninos, T. Anous, and F. Denef, \textit{%
Disordered quivers and cold horizons}, J. High Energy Phys. 12, 071 (2016).

\bibitem{Weidenmuller} L. Benet, T. Rupp and H. A. Weidenmüller, Ann. Phys. 292, 67 (2001);
T. Asaga, L. Benet, T. Rupp and H. A. Weidenmüller, Ann. Phys. 298, 229 (2002).

\bibitem{1908.00995} Jan Behrends and Benjamin Beri, \textit{Supersymmetry
in the nonsupersymmetric Sachdev-Ye-Kitaev model}, preprint arXiv:1908.00995.

\bibitem{1912.09975} Jan Behrends and Benjamin Beri, \textit{Symmetry
classes, many-body zero modes, and supersymmetry in the complex
Sachdev-Ye-Kitaev model}, preprint arXiv:1912.09975.

\bibitem{Debanjan} Debanjan Chowdhury and Erez Berg, \textit{Intrinsic superconducting instabilities of a solvable model for an incoherent metal},
Phys. Rev. Research 2, 013301 (2020).

\bibitem{Murugan1} Jeff Murugan, Douglas Stanford, and Edward Witten,
\textit{More on supersymmetric and 2d analogs of the SYK model}, J. High
Energ. Phys. 2017, 146 (2017).

\bibitem{Murugan2} J. Murugan and H. Nastase, \textit{One-dimensional
bosonization and the SYK model}, J. High Energ. Phys. 2019, 117 (2019).

\bibitem{Yevtushenko-2003} O. Yevtushenko and V. E. Kravtsov, \textit{Virial
expansion for almost diagonal random matrices}, J. Phys. A 36, 8265 (2003).

\bibitem{Yevtushenko-2004} O. Yevtushenko and V. E. Kravtsov, \textit{%
Density of states for almost-diagonal random matrices}, Phys. Rev. E 69,
026104 (2004).

\bibitem{oleg1} O. Yevtushenko and A. Ossipov, \textit{A supersymmetry
approach to almost diagonal random matrices}, Journal of Physics A:
Mathematical and Theoretical 40 (18), 4691 (2007).

\bibitem{oleg2} S. Kronmueller, O. M, Yevtushenko, and E. Cuevas, \textit{%
Supersymmetric virial expansion for time-reversal invariant disordered
systems}, Journal of Physics A: Mathematical and Theoretical 43 (7), 075001
(2010).

\bibitem{Efetov-2003} K. B. Efetov and V. R. Kogan, \textit{Nonlinear $%
\sigma $ model for long-range disorder and quantum chaos}, Phys. Rev. B 67,
245312 (2003).

\bibitem{Efetov-2004} K. B. Efetov, G. Schwiete, and K. Takahashi, \textit{%
Bosonization for Disordered and Chaotic Systems}, Phys. Rev. Lett. 92,
026807 (2004).

\bibitem{Sedrakyan-2005} T. Sedrakyan, \textit{Toda lattice representation
for random matrix model with logarithmic confinement}, Nucl. Phys. B 729,
526 (2005).

\bibitem{Sedrakyan-2017} T. A. Sedrakyan, and K. B. Efetov, \textit{%
Superbosonization in disorder and chaos: Role of anomalies}, Phys. Rev. B
96, 054208 (2017).

\bibitem{berezin} F. A. Berezin, \textit{Introduction to Superanalysis},
MPAM Vol. 9 (Reidel, Dordrecht, 1987).

\bibitem{km} A. Kamenev and M. Mezard, \textit{Wigner-Dyson statistics from
the replica method}, J. Phys. A: Math. Gen. 32 4373 (1999); \textit{Level
correlations in disordered metals: The replica $\sigma$ model}, Phys. Rev. B
60, 3944 (1999).

\bibitem{kanzieper} E. Kanzieper, \textit{Replica Field Theories, Painleve
Transcendents, and Exact Correlation Functions} Phys. Rev. Lett. 89, 250201
(2002); K. Splittorff and J. J. M. Verbaarschot, \textit{Replica Limit of
the Toda Lattice Equation}, Phys. Rev. Lett. 90, 041601 (2003).



\bibitem{Altshuler-1988} B. Altshuler, I. Zarekeshev, S. Kotochigova, and B.
Shklovskii, \textit{\ Repulsion between energy levels and the
metal-insulator transition}, Sov. Phys. JETP [Zh. Eksp. Teor. Fiz. 94, 343]
67, 15 (1988).

\bibitem{Braun-1995} D. Braun and G. Montambaux, \textit{Spectral
correlations from the metal to the mobility edge}, Phys. Rev. B 52, 13903
(1995).

\bibitem{Bertrand-2016} C. L. Bertrand and A. M. Garcia-Garcia, \textit{%
Anomalous Thouless energy and critical statistics on the metallic side of
the many-body localization transition}, Phys. Rev. B 94, 144201 (2016).

\bibitem{Benet-2001} L. Benet, T. Rupp, and H. A. Weidenmuller, \textit{%
Nonuniversal Behavior of the $k$-Body Embedded Gaussian Unitary Ensemble of
Random Matrices}, Phys. Rev. Lett. 87, 010601 (2001).


\bibitem{moura} F.A.B.F. de Moura and Marcelo L. Lyra, \textit{%
Delocalization in the 1D Anderson Model with Long-Range Correlated Disorder}%
, Phys. Rev. Lett. 81, 3735 (1998).

\bibitem{Izrail} F. M. Izrailev and A. A. Krokhin, \textit{Localization and
the Mobility Edge in One-Dimensional Potentials with Correlated Disorder},
Phys. Rev. Lett. 82, 4062 (1999).

\bibitem{Garcia} A. M. Garcia-Garcia and E. Cuevas, \textit{Differentiable
potentials and metallic states in disordered one-dimensional systems}, Phys.
Rev. B 79, 073104 (2009).

\bibitem{SDS} X. C. Xie and S. Das Sarma, \textit{"Extended" electronic
states in a Fibonacci superlattice}, Phys. Rev. Lett. 60, 1585 (1988).

\bibitem{Flores} J. C. Flores, \textit{Transport in models with correlated
diagonal and off-diagonal disorder}, J. Phys. Condens. Matter 1, 8471 (1989).

\bibitem{Dunlap} D. H. Dunlap, H.-L. Wu, and P. Phillips, \textit{Absence of
localization in a random-dimer model}, Phys. Rev. Lett. 65, 88 (1990).

\bibitem{PP} P. Phillips and H.-L. Wu, \textit{Localization and Its Absence:
A New Metallic State for Conducting Polymers}, Science 252, 1805 (1991).


\bibitem{TS1} T. Sedrakyan, \textit{Localization-delocalization transition
in a presence of correlated disorder: The random dimer model}, Phys. Rev. B
69, 085109 (2004).

\bibitem{TS2} T. Sedrakyan and A. Ossipov, \textit{%
Localization-delocalization transition in the quasi-one-dimensional ladder
chain with correlated disorder}, Phys. Rev. B 70, 214206 (2004).

\bibitem{TS3} T. A. Sedrakyan, J. P. Kestner, S. Das Sarma, \textit{Proposed
signature of Anderson localization and correlation-induced delocalization in
an N-leg optical lattice}, Phys. Rev. A 84, 053621 (2011).

\bibitem{Zhuang-2017} Yi-Zhuang You, Andreas W. W. Ludwig, and Cenke Xu,
\textit{Sachdev-Ye-Kitaev model and thermalization on the boundary of
many-body localized fermionic symmetry-protected topological states}, Phys.
Rev. B 95, 115150 (2017).

\bibitem{Li-2017} Tianlin Li, Junyu Liu, Yuan Xin, and Yehao Zhou, \textit{%
Supersymmetric SYK model and random matrix theory}, Journal of High Energy
Physics, 2017(6), 111 (2017).

\bibitem{Hunter-Jones-2017} N. Hunter-Jones and J. Liu, \textit{Chaos and
random matrices in supersymmetric SYK}, JHEP 1805, 202 (2018).

\bibitem{Cotler-2017-2} Jordan Cotler, Nicholas Hunter-Jones, Junyu Liu, and
Beni Yoshida, \textit{Chaos, complexity, and random matrices}, Journal of
High Energy Physics, 2017(11), 48 (2017).

\bibitem{delCampo-2017} A. del Campo, J. Molina-Vilaplana, and J. Sonner,
\textit{Scrambling the spectral form factor: Unitarity constraints and exact
results}, Phys. Rev. D, 95, 126008 (2017).

\bibitem{Miklitz} T. Micklitz, Felipe Monteiro, and Alexander Altland,
\textit{Nonergodic Extended States in the Sachdev-Ye-Kitaev Model}, Phys.
Rev. Lett. 123, 125701 (2019).

\bibitem{footnote} Note that we used the decoupling with two
Hubbard-Stratonovich fields in Sections III and IV as the saddle point
equations following from a single field decoupling there did not produce the
scheme of Ref.~\onlinecite{Sachdev-2015} for the calculation of the Green's
function.

\bibitem{Jia-2019} Yiyang Jia and J.M. Verbaarshot,
\textit{Spectral Fluctuations in the Sachdev-Ye-Kitaev Model}, arXiv:1912.11923v3.

\bibitem{Elitzur-1986} S. Elitzur,  E Rabinovici, Y. Frishman and 
A. Schwimmer, 
\textit{Origins of Global Anomalies in Quantum Mechanics}, Nucl. Phys.  B273. 93 (1986).


\end{thebibliography}
\end{document}